\documentclass[aps,preprintnumbers,amsmath,amssymb,nofootinbib]{revtex4}
\usepackage[T1]{fontenc}
\usepackage{enumitem}
\usepackage{graphicx}
\usepackage{hyperref}
\hypersetup{colorlinks=true,urlcolor=blue,linkcolor=black,citecolor=black}
\usepackage{amsmath,amssymb}
\usepackage{xfrac}
\usepackage{amsthm}

\newcommand{\Ha}[1]{\mathcal{H}\left[ #1\right]}
\newcommand{\M}[1]{\mathcal{M}\left[ #1\right]}
\newcommand{\Echi}[1]{\mathbb{E}_{\chi}\left[ #1\right]}
\newcommand{\Sobol}[1]{\mathcal{H}_{L}\left[ #1\right]}

\newcommand{\epsfNLSref}{\hyperref[eq:EpsfNLS]{($\varepsilon$-fNLS)}}

\newcommand{\Psinu}{\psi_\nu}
\newcommand{\Psisig}{\psi_\sigma}
\newcommand{\Psieps}{\psi_{\varepsilon}}

\newcommand{\T}{\mathbb{T}}
\newcommand{\R}{\mathbb{R}}

\newtheorem{proposition}{Proposition}

\newtheorem*{remark*}{Remark}

\begin{document}

\title{Wave collapse proliferation as a mechanism for anomalous dissipation and spontaneous stochasticity}

\author{Wandrille Ruffenach}

\email{wandrille.ruffenach@ens-lyon.fr}

\affiliation{ENS de Lyon, CNRS, LPENSL, UMR5672, 69342, Lyon cedex 07, France.}

\date{November 13, 2026}
\begin{abstract}
Anomalous dissipation and spontaneous stochasticity are signatures of the roughness of the inviscid limit in fluid turbulence. Spontaneous stochasticity describes the selection of a probability distribution over possible solutions rather than a single deterministic trajectory. We investigate these phenomena in a nonlinear dispersive equation undergoing finite-time blowup, known as wave collapse. Collapse is arrested by two regularization mechanisms, one dissipative and one conservative. Heuristic arguments supported by numerical simulations show that an isolated collapse has a vanishing effect in the inviscid limit and cannot alone account for either phenomenon. However, simulations reveal that the number of collapse events diverges as the regularization vanishes. Their proliferation enables sustained dissipation in the dissipative case and prevents either regularization from selecting a single inviscid solution. We then introduce vanishingly small random perturbations and find numerical evidence for a well-defined statistical limit in which randomness persists. These results identify collapse proliferation as a mechanism connecting anomalous dissipation and spontaneous stochasticity in dispersive waves, allowing a formally deterministic equation to acquire a truly random limiting evolution.
\end{abstract}


\maketitle
\section{Introduction}
\label{sec:introduction}

Predictability in deterministic systems is often limited by chaos. Nearby trajectories separate exponentially in time, and the largest Lyapunov exponent sets a predictability horizon. Yet chaos does not by itself destroy determinism. If the initial condition were known with infinite precision, the future state would still be uniquely determined, and randomness would only reflect incomplete information. Some systems fall outside this paradigm and display a loss of predictability in finite time. This phenomenon is commonly referred to as spontaneous stochasticity.

A standard framework for discussing spontaneous stochasticity starts from a deterministic but ill-posed inviscid model, such as the Euler equations, which admits several weak solutions. In practice, such inviscid systems are not observed directly: they are regularized by additional physical effects, for instance viscous diffusion in fluid flows. When the regularization is sent to zero, the regularized dynamics might either select a unique solution of the inviscid model or fail to converge among its possible different solutions. In the latter case, the regularization does not provide a deterministic selection principle, and the inviscid limit is more naturally described in statistical terms. This is the basic mechanism underlying spontaneous stochasticity.

Spontaneous stochasticity has been investigated in a variety of fluid systems, including Navier-Stokes turbulence \cite{ThaBec20,BifBof18}, passive scalar advection \cite{RufSim25}, surface quasi-geostrophic flows \cite{PalDor14,RotSny08,ValTha24}, and reduced models of turbulence \cite{CriJen93,Mai16,Mai16b,BanMai24,Mai24,OrtCam26,PeiTha23}. Proving that a system is spontaneously stochastic in the sense of \cite{RufSim26} remains difficult and depends sensitively on both the inviscid model and the regularization mechanism. Finite-dimensional systems have proven especially useful in this respect, since they often allow analytical predictions \cite{RufSim26,DriMai21,DriMai24,BarChe26}, sometimes with the help of renormalization-group methods \cite{RufSim26,Mai24,MaiRai23,EyiBan20,Mai26}. To the best of our knowledge, the work \cite{RufSim25} on the passive scalar model of \cite{ArmVic25} remains the only rigorous proof of spontaneous stochasticity for the Cauchy problem of a partial differential equation, that is, of Eulerian spontaneous stochasticity. See also \cite{FlaReh24} for ideas relating spontaneous stochasticity and convex integration for Euler equations.

All in all, spontaneous stochasticity may be regarded as one of the characteristic features of fluid turbulence \cite{RufSim25,EyiGol25} and motivates stochastic modeling \cite{PerGar16,ChaLem26,BecBre24}.

Turbulent behavior is not restricted to fluid systems. It also occurs in nonlinear dispersive media, where complexity is governed by the interplay of dispersion and nonlinearity. A paradigmatic example is provided by the nonlinear Schr\"odinger equation (NLS) \cite{Naz11} and by its one-dimensional nonlocal counterpart, the Majda-McLaughlin-Tabak (MMT) model of wave turbulence \cite{MajMcL97,CaiMaj01}. In the focusing regime, NLS-type equations may undergo wave collapse, namely finite-time concentration of the field into a singular region. During collapse, the amplitude diverges in finite time while conserved quantities such as the mass remain finite. The theory of continuation past collapse includes nonunique solutions, regularization-dependent mass loss and loss of the phase of the collapsing core \cite{Mer92a,Tao09,FibKle11,FibKle12}. These results make collapsing dispersive equations natural candidates for investigating whether phenomena akin to spontaneous stochasticity can arise outside the standard hydrodynamic setting. Nonlinear optical media may also offer an experimental testbed: phase loss in collapsing beams has been observed and its influence on later interactions has been studied \cite{ShiSch12,SagDit17}, revealing a possible route toward spontaneous stochasticity.

In this work, we study spontaneous stochasticity in a collapsing fractional nonlinear Schr\"odinger model. After recalling the general framework of spontaneous stochasticity, we introduce the inviscid model and two different collapse-arresting regularizations. The first one is a viscous diffusion, motivated by fluid turbulence, and the second is a saturating nonlinearity, motivated by nonlinear optics. For a fixed regularization parameter, both mechanisms prevent blowup. When the regularization is removed, however, both recover finite-time collapse. The corresponding inviscid limits display both common and mechanism-dependent features. Before collapse, they converge toward the same smooth inviscid solution. After collapse, however, the solutions retain regularization dependent features. The viscous approximation displays a finite loss of mass in the vanishing-viscosity limit, giving a dispersive analogue of anomalous dissipation already observed in related models \cite{JosPom20,AmaJos23}. By contrast, the saturating approximation remains conservative.

The point developed here is that collapse is weak: the mass contained in a shrinking core tends to zero, as in mass-supercritical NLS collapse \cite{SulSul07,Fib15,MerRapSze10}. An isolated event can therefore only have a vanishing effect on both total mass and field uncertainty. A finite effect requires a collective mechanism, as suggested by the singularity-turbulence scenario \cite{JosPom20,AmaJos23}. We demonstrate that the proliferation of collapse cores, related to the collapse-turbulence regimes reported for MMT systems \cite{RumBiv05,RumShe15,SimPan24}, and their subsequent interactions compensate for the decreasing mass affected by each event. Starting from a smooth localized initial condition, numerical integrations of the regularized dynamics show that the number of collapse events diverges in the inviscid limit, while the dissipation and uncertainty produced by each event vanish. We then investigate spontaneous stochasticity: can randomness of vanishing amplitude persist in this limit \cite{RufSim25,RufSim26}? Specifically, we perturb either the initial condition, with an amplitude that vanishes in the inviscid limit, or the regularization parameter.

Our results show that collapsing fractional nonlinear Schr\"odinger equation provides a one-dimensional dispersive setting in which singular inviscid limits, anomalous dissipation, and spontaneous stochasticity can be studied within a common framework. In addition to being a conceptually useful toy model for fluid turbulence, this also makes nonlinear optical systems natural candidates for possible experimental work on spontaneous stochasticity.

\section{The spontaneous stochasticity triptych}
\label{sec:SpTrip}

The present work is organized around the measure selection viewpoint of spontaneous stochasticity formalized and developed in \cite{RufSim26} that we recall here for the sake of clarity. We present only the ingredients needed for our purposes and refer to \cite{RufSim26} for the rigorous measure-theoretic formulation and further results. The central point is that spontaneous stochasticity is a measure-selection principle for singular deterministic limits: when a vanishing regularization does not select a single inviscid state, it may still select a probability law on the set of possible inviscid states. This viewpoint can be organized into the following triptych.

\begin{enumerate}[label=\textbf{(T\arabic*)},ref=T\arabic*]
\item \label{it:spst-inviscid} \textbf{Ill-posed inviscid dynamics.} One starts from an inviscid, or unregularized, system
\begin{equation}
(\mathcal P_0) \qquad \dot x = f_0(x), \qquad x(0)=x_0. \label{eq:abstract-inviscid}
\end{equation}
Typically, the vector field $f_0$ is not Lipschitz, leading to possibly infinitely many solutions of $(\mathcal P_0)$. In a fluid-turbulence context, one may think of \eqref{eq:abstract-inviscid} as the Euler equation in a regime where the solution is very rough. This nonuniqueness should not be interpreted as something directly observed in nature. Physical systems do not realize the inviscid equation in isolation. They contain small-scale effects, such as viscosity, dispersion, damping, or microscopic cutoffs, which act as regularization mechanisms for the dynamics. The inviscid problem is therefore probed only indirectly, through a family of regularized problems. The relevant question is not merely whether \eqref{eq:abstract-inviscid} has several solutions, but which of these solutions, if any, are selected by physically meaningful approximations.
\item \label{it:spst-regularized} \textbf{Well-posed regularized dynamics.} One therefore introduces a family of regularized problems
\begin{equation}
(\mathcal P_\varepsilon) \qquad \dot x_\varepsilon = f_\varepsilon(x_\varepsilon), \qquad x_\varepsilon(0)=x_0, \label{eq:abstract-regularized}
\end{equation}
where $\varepsilon>0$ is a small parameter controlling the regularization and $f_\varepsilon\to f_0$ as $\varepsilon\to 0$. The regularization is assumed to restore well-posedness for every fixed $\varepsilon>0$. In the fluid analogy, this is the role played by Navier-Stokes as the vanishing-viscosity approximation of Euler. The selection question is then: does the regularized dynamics $(\mathcal P_\varepsilon)$ select a unique solution of $(\mathcal P_0)$ in the inviscid limit $\varepsilon\to 0$? If for a given $t>0$, $x^\varepsilon(t)$ has no limit when $\varepsilon \to 0$, the regularization fails to provide a deterministic selection principle. In that case, the vanishing-regularization procedure does not allow one to deterministically select a trajectory of the inviscid system.

\item \label{it:spst-statistical} \textbf{Vanishing ambient randomness.} The deterministic inviscid limit considered in \ref{it:spst-regularized} can fail in two distinct ways. First, for a fixed initial condition, the regularized states $x^\varepsilon(t)$ may fail to converge as $\varepsilon \to 0$, so that the regularization does not select a unique inviscid state. Second, continuity with respect to the initial condition may be lost in the singular limit: perturbations of the initial datum that vanish together with the regularization may still produce an order-one discrepancy at finite time. In each case, an ambient randomness on either the regularization parameter and/or on the initial condition can survive in the joint vanishing regularization and randomness limit. If randomness persists in the limit, it might be possible to give a well-defined probabilistic meaning to the inviscid limit. We recall briefly how this is done in \cite{RufSim26} when randomness is introduced on the regularization parameter and refer to \cite{RufSim26} for the discussion on the initial condition. Let the regularization parameter $\varepsilon$ be sampled according to a probability measure $\mathbb P_\eta$ concentrated near zero (typically with support on $[0,\eta]$), and define, at fixed time $t$, $\gamma:\varepsilon\longmapsto x^\varepsilon(t)$ encoding the dependence of the solution of \eqref{eq:abstract-regularized} on the regularization parameter. For a random regularization parameter $\varepsilon(\omega)$, the random state $x^{\varepsilon(\omega)}(t)$ is distributed according to the so-called pushforward measure $\mu_\eta=\gamma_\#\mathbb P_\eta$. Since the purpose is to probe the deterministic inviscid problem $(\mathcal P_0)$, the ambient randomness must disappear together with the regularization, namely $\mathbb P_\eta \rightharpoonup \delta_0$ as $\eta\to 0$. The limit $\eta\to 0$ is then simultaneously a vanishing-regularization and vanishing-randomness limit, and the object of interest is the limiting behavior of the induced law $\mu_\eta$. An analogous construction can be performed by randomizing a family of initial conditions converging to the prescribed datum; we refer to \cite{RufSim26} for a detailed treatment of this case.
\end{enumerate}

Once the triptych \ref{it:spst-inviscid}-\ref{it:spst-statistical} is fixed, spontaneous stochasticity is defined through the limiting behavior of $\mu_\eta$. If $\mu_\eta \rightharpoonup \delta_{\tilde x}$, for some inviscid state $\tilde x$, then the limiting procedure selects a deterministic state. If instead $\mu_\eta \rightharpoonup\mu$, with $\mu$ non-Dirac, then a vanishing microscopic randomness produces a macroscopic uncertainty at finite time, and the inviscid limit is spontaneously stochastic. Finally, if $\mu_\eta$ has several subsequential limits, then even the limiting statistics are not uniquely selected. Following \cite{RufSim26}, these three possibilities correspond, respectively, to deterministic selection, strong spontaneous stochasticity, and weak spontaneous stochasticity, up to the pathological Dirac-limit cases discussed in \cite{RufSim26}. As detailed in \cite{RufSim26}, the weak/strong dichotomy arises from a subtle interplay between the regularization and the ambient measure.

The present paper is built around this triptych. Section~\ref{sec:InviscidModel} introduces the inviscid model and identifies wave collapse as the singular event leading to non uniqueness, that is \ref{it:spst-inviscid}. Section~\ref{sec:REGStopCollapse} introduces two collapse-arresting mechanisms, viscous diffusion and nonlinear saturation, which implement \ref{it:spst-regularized}. In Section~\ref{sec:BowupAD} we examine whether both regularizations recover inviscid collapse in the limit. Then, we show that the two mechanisms lead to distinct post-blowup limits, one dissipative and one conservative, giving a first indication that there exist multiple post blowup continuations. Although leading to qualitatively different inviscid behavior, each regularization mechanism could act as selection principle and therefore rule out spontaneous stochasticity. In Section~\ref{sec:LSP_reg} we show that it is not the case, for each regularization mechanism the vanishing regularization state is wandering in the space of inviscid solutions without converging. Additionally, in Section~\ref{sec:LCWRTIC}, we show that vanishingly small perturbations of the initial condition also lead to finite time unpredictability of the limit. These observations points toward a statistical description of the inviscid limit rather than a deterministic one, that is spontaneous stochasticity. Finally, Section~\ref{sec:StatInvLimit} we study the statistical behavior of the limit . In accordance with \ref{it:spst-statistical}, this is done by introducing a source of randomness on either the regularization parameter or on the initial condition. All in all, we test two regularizations and two different sources of randomness and explore the associated inviscid statistics. The numerical results show that, as expected, the pre-blowup state is deterministic, whereas after blowup it remains genuinely random. We compute the probability distributions induced by each source of randomness and show that, for a randomized initial condition, the distribution of the Fourier modes converges toward a well-defined law independent of the regularization parameter, while for a randomized regularization parameter it shows no clear sign of convergence. This illustrates both the weak and the strong forms of spontaneous stochasticity discussed above.

\section{The inviscid model}
\label{sec:InviscidModel}

Finite-time blowup in focusing nonlinear Schr\"odinger-type equations results from the competition between two effects: focusing nonlinearity, which tends to concentrate the solution at a point, and dispersion, which tends to spread it. We consider the focusing fractional nonlinear Schr\"odinger equation
\begin{equation}
\label{eq:MMT} \tag{fNLS}
\begin{cases}
i\partial_t \psi = \Lambda^\alpha \psi -|\psi|^{2} \psi, & (x,t) \in \T\times \R^+, \\
\psi|_{t=0}=\psi_0 \in \mathcal{C}^\infty(\T),
\end{cases}
\end{equation}
where $\T=\R /(2\pi \mathbb Z) $ is the one dimensional torus and $\Lambda^{\alpha}$ is the Fourier multiplier $|k|^{\alpha}$ with Fourier modes defined as
$$
\widehat{\psi}(k)= \dfrac{1}{2\pi} \int_\T \psi(x) e^{-ikx}dx, \qquad \psi(x) = \sum_{k \in \mathbb{Z}} \widehat{\psi}(k) e^{ikx}.
$$
The inviscid dynamics \eqref{eq:MMT} formally conserves the mass $\mathcal{M}$ and the Hamiltonian $\mathcal{H}$,
\begin{align}
&\label{eq:Mass} \M{\psi}=\int_{\T} |\psi(t,x)|^2 \, d x,\\
&\label{eq:Hamil} \Ha{\psi}=\int_{\T} \dfrac{1}{2}\left|\Lambda^{\frac \alpha 2} \psi(t,x)\right|^2 -\dfrac{1}{4}|\psi(t,x)|^{4} \, d x.
\end{align}

In the focusing case considered here, where the cubic nonlinearity enters with a negative sign, solutions of \eqref{eq:MMT} may undergo wave collapse at a finite time $0<t^\star<+\infty$, in the sense that
\begin{equation}
\label{eq:BlowupCrit} \lim_{t \rightarrow t^\star} \|\Lambda^{\frac{\alpha}{2}} \psi(t,\cdot)\|_{L^2(\T) } = + \infty.
\end{equation}
Here the $L^2$ norm $\|\cdot\|_{L^2(\T)}$ is the square root of the mass \eqref{eq:Mass}. Such finite time blowup is usually referred to as wave collapse since in practice, the solution concentrates around a point and its amplitude diverges. By denoting $\|\psi(t,\cdot)\|_\infty= \sup_{x\in\T} |\psi(t,x)|$ one indeed has
$$
\|\Lambda^{\frac{\alpha}{2}} \psi(t,\cdot)\|^2_{L^2(\T) } =2 \Ha{\psi_0} + \int_\T \dfrac{1}{2}|\psi(t,x)|^{4} \, d x\leq2 \Ha{\psi_0} + \dfrac{1}{2}\M{\psi_0} \|\psi(t,\cdot)\|_\infty^2.
$$
Thus, at the blowup time $t^\star$, both the field amplitude and the mass of its fractional derivative of order $\alpha/2$ diverge.

We use \eqref{eq:BlowupCrit} as the collapse criterion for the solution studied here. Its relevance and relation to mass and energy criticality is discussed below. For such finite-time blowup to occur, nonlinear effects must dominate (remain at least comparable to) dispersive ones near the collapsing core. This can be understood with the following heuristic argument. Consider a localized structure, or bump, of width $L(t)$ and fixed unit mass. In spatial dimension $d$ and with a nonlinearity $|\psi|^{2p}\psi$, dimensional analysis gives $|\psi|\sim L^{-d/2}$ because the solution has unit mass. The dispersive timescale then scales as $\tau_d\sim L^\alpha$, whereas the nonlinear timescale scales as $\tau_{\rm NL}\sim |\psi|^{-2p}\sim L^{pd}$. Their ratio behaves as
\[
\frac{\tau_d}{\tau_{\rm NL}} \propto L^{\alpha-pd}.
\]
Finite time wave collapse, meaning that $L(t)\to0$ as $t\to t^\star$, can occur only if the nonlinear timescale remains smaller, or at least compares to, the dispersive timescale during concentration. Otherwise dispersion would scatter the core away. Thus the above ratio must not vanish as $L\to0$, which gives the criterion $ \alpha\leq pd$ for finite time wave collapse to occur.

In the classical case $\alpha=2$, this condition recovers the usual classification of NLS dynamics: the mass-critical regime $p d = 2$ and the mass-supercritical regime $p d > 2$, in which finite-time blowup can occur \cite{SulSul07}. In the mass-critical case, the ground state sets the mass threshold, while stable self-similar blowup is known for slightly mass-supercritical equations \cite{SulSul07,Fib15,MerRapSze10}. Consequently, standard cubic NLS ($p=1$) can exhibit collapse only in dimensions $d \geq 2$. In contrast, the fractional cubic NLS permits collapse already in one spatial dimension provided $\alpha \leq 1$, which makes it particularly attractive for numerical investigations.

A physically relevant case is $\alpha = 1/2$, for which the square-root dispersion is that of deep-water gravity waves. It is also the standard choice in the focusing Majda-McLaughlin-Tabak family of wave models \cite{MajMcL97,CaiMaj01,RumBiv05,RumShe15,ChiDeL17}, and we adopt it here.

In the above heuristic argument, we supposed that the bump of width $L$ carried one unit of mass. An actual collapsing core need not do so. Let $A(t)$ denote its amplitude. Balancing the nonlinear and dispersive frequencies gives

\begin{equation}
\label{eq:WeakCore} A\sim L^{-\alpha/(2p)},\qquad m_{\rm core}\sim A^2L^d\sim L^{d-\alpha/p}.
\end{equation}
Thus the core carries finite mass in the mass-critical case $\alpha=pd$, but vanishing mass in the mass-supercritical case $\alpha<pd$. This regime is known as weak collapse \cite{ZakLitRak88,SulSul07,Fib15}. The vanishing mass of each core is central to the proliferation mechanism studied below: a finite aggregate effect cannot be produced by a finite number of events.

The same argument can be performed with the assumption of a unit Hamiltonian rather than mass. This leads to the energy sub-critical or super-critical regimes. Indeed, under the balance \eqref{eq:WeakCore}, the kinetic and nonlinear contributions scale in the same way,

\[
\|\Lambda^{\alpha/2}\psi\|_{L^2}^2 \sim \|\psi\|_{L^{2p+2}}^{2p+2} \sim L^{d-\alpha\frac{p+1}{p}}.
\]
Their individual contributions therefore diverge in the energy-subcritical case $\alpha>pd/(p+1)$, while the Hamiltonian remains constant. In the energy-supercritical case $\alpha<pd/(p+1)$, a collapsing core may instead carry a vanishing contribution to both terms. At energy criticality, $\alpha=pd/(p+1)$, the leading exponent is zero and this dimensional argument decides neither whether the critical Sobolev norm remains bounded nor whether it grows slowly. Accordingly, \eqref{eq:BlowupCrit} is the relevant diagnostic for the present solution, but it is not a universal characterization of collapse in energy-critical or energy-supercritical problems. Classical energy-critical NLS, for example, admits finite-time type-II collapse with bounded critical Sobolev norm \cite{Schmid23}.

Notably, the case $\alpha = 1/2$ is both mass-supercritical and energy-critical, the latter being unattainable for the standard ($\alpha = 2$) one-dimensional NLS with general nonlinearity \cite{SulSul07,Fib15}. Together with the nonlocal dispersion, this is an important difference from the usual one-dimensional focusing NLS. For the collapse studied here, we will show numerically that the critical Sobolev norm $\|\Lambda^{1/4}\psi(t)\|_{L^2}^2$ grows logarithmically as $t\to t^\star$. This divergence being slower than any power law, it is consistent with the heuristic power counting argument above that capture only power law divergence.

Rigorous virial arguments establish blowup for fractional NLS on $\R^d$ and on bounded domains in other exponent ranges \cite{BouHimLen16}, but they do not cover the periodic cubic equation at $\alpha=1/2$. For one-dimensional fractional NLS, numerical evidence supports collapse and the leading self-similar rates in mass-critical and mass-supercritical regimes \cite{KleSpa14}; the focusing energy-critical case considered here was not resolved there.

If the contraction time is also comparable to the dispersive and nonlinear times, then $t^\star-t\sim L^\alpha$. For the cubic equation, this gives the leading powers

\begin{equation}
\label{eq:WeakTimeScaling} L(t)\sim(t^\star-t)^{1/\alpha},\qquad \|\psi(t)\|_\infty\sim(t^\star-t)^{-1/2},\qquad \dot\gamma(t)\sim(t^\star-t)^{-1},
\end{equation}
where $\gamma$ is the phase of the core. At energy criticality, the length retains this leading scaling, but the amplitude may carry a slowly varying correction. That is for example the case of mass-critical collapse in classical NLS which carries a log-log correction \cite{Fib15,SulSul07}. Given the resolution of our numerical simulations and the scope of the present article, we will not try here to describe quantitatively these possible sub-power law corrections. We note however that they are most likely present since the critical Sobolev norm diverges logarithmically. In the following, we will use extensively the heuristic scaling presented here and compare it favorably to our numerical results in the range of attainable resolutions.

The dynamics \eqref{eq:MMT} therefore constitutes the inviscid model of interest for the first part \ref{it:spst-inviscid} of the spontaneous stochasticity triptych. The smooth Cauchy problem is well posed up to $t^\star$, but collapse alone proves neither existence nor nonuniqueness of a weak continuation. To our knowledge, no global post-collapse continuation theory is available for the periodic fractional equation considered here. Section~\ref{sec:REGStopCollapse} therefore first reviews the corresponding results for the classical NLS and then introduces a collapse-arresting mechanism.

\section{Regularization of wave collapse}
\label{sec:REGStopCollapse}

Before $t^\star$, \eqref{eq:MMT} admits a unique smooth solution. After collapse, however, the notion of solution must be weakened and its continuation can depend on how the singularity is arrested. Merle constructed nonunique continuations associated with critical-mass NLS singular solutions \cite{Mer92a}. Tao constructed global weak solutions for mass-critical NLS in $d\geq4$ and also identified a more restrictive continuation class in which uniqueness is recovered \cite{Tao09}.

Several regularizations have been investigated in the NLS literature. Fibich and Klein compared sub-threshold, shrinking-hole, damping and complex Ginzburg-Landau continuations and showed that different vanishing sequences can leave different phases after collapse \cite{FibKle11}. Nonlinear damping can produce a strongly asymmetric continuation with rapid expansion after arrest \cite{FibKle12}. Phase loss has also been observed in collapsing optical beams \cite{ShiSch12}; its consequences for long-distance stochastic beam interactions were studied in \cite{SagDit17}. That long-distance randomization limit differs from the fixed-time, vanishing-perturbation limit considered here.

\subsection{Regularized dynamics}

To study the post-blowup dynamics in a controlled way, a standard strategy is to regularize the equation, obtain a unique global-in-time strong solution for each fixed value of the regularization parameter, and then consider the vanishing-regularization limit. If this limit fails to select a unique weak solution of the inviscid problem \eqref{eq:MMT}, the regularization mechanism provides a natural framework for spontaneous stochasticity as emphasized in Section~\ref{sec:SpTrip}.

\begin{figure}[t]
\centering
\includegraphics[width=0.97\linewidth]{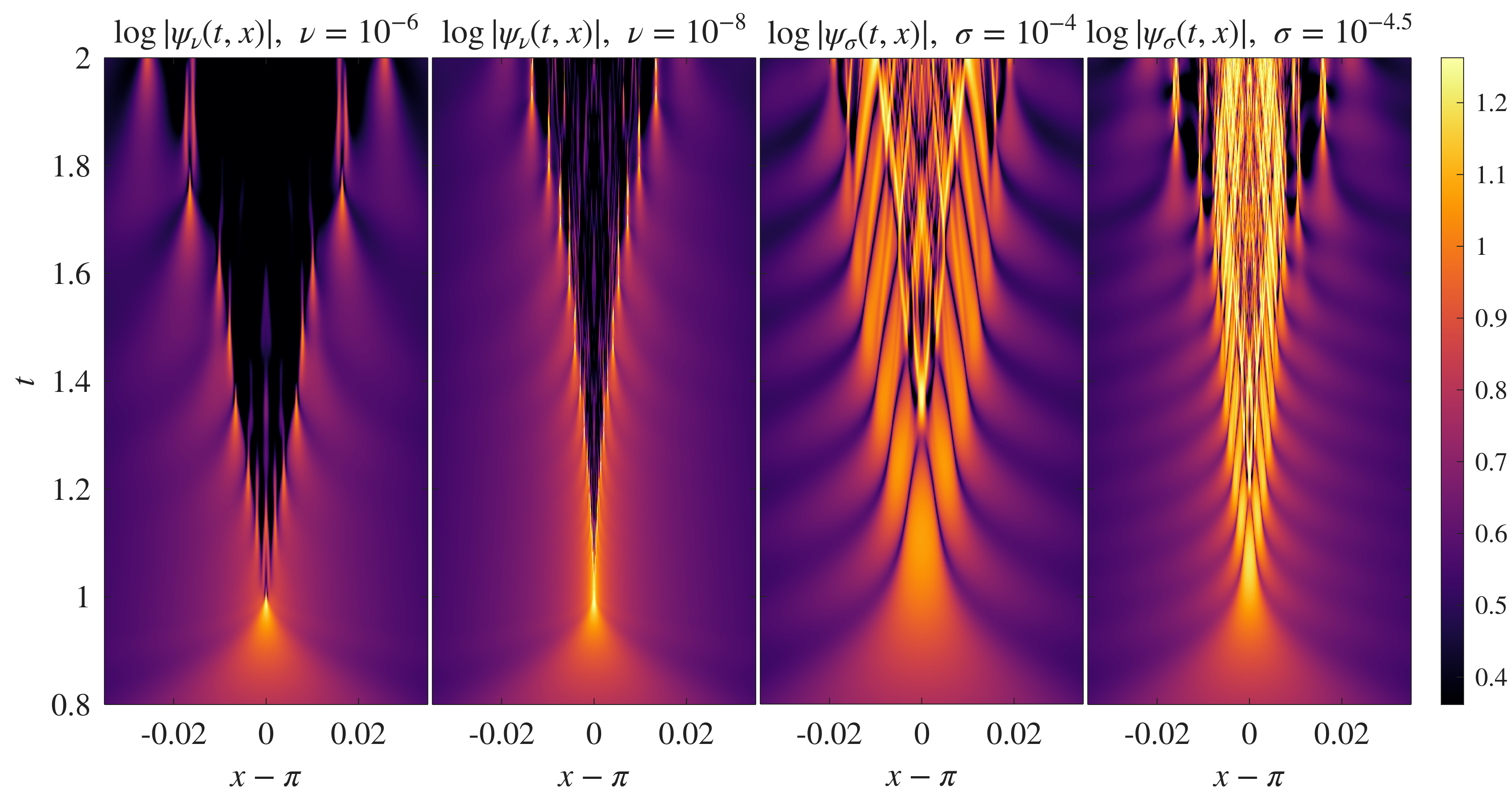}
\caption{Space-time diagrams of the amplitude of the solution for both regularizations, \eqref{eq:ViscousMMT} and \eqref{eq:SaturMMT}, at two levels of regularization each. The value of the regularization parameter is given in the title of each panel. For illustrative movies of the results presented in this paper, see \href{https://youtube.com/playlist?list=PLYfX0VUIlSSk}{these videos}.}
\label{fig:SpaceTimeIllustr}
\end{figure}

In the present work, we focus on two regularizations of the inviscid dynamics \eqref{eq:MMT}. The first is a viscous diffusion term, commonly used as small-scale damping in wave-turbulence simulations of the MMT model, often in the form of hyperviscosity. This choice is motivated by the analogy with fluid turbulence. The second regularization is a saturating nonlinearity, closer to nonlinear optics. The corresponding regularized equations that we will study throughout this paper read
\begin{subequations}
\label{eq:EpsfNLS}
\begin{align}
i\partial_t \Psinu &= \Lambda^\alpha \Psinu -|\Psinu|^2 \Psinu + i\nu\Delta \Psinu, \tag{$\nu$-fNLS}\label{eq:ViscousMMT} \\[0.3em]
i\partial_t \Psisig &= \Lambda^\alpha \Psisig -\frac{|\Psisig|^2}{1+\sigma|\Psisig|^4}\,\Psisig, \tag{$\sigma$-fNLS}\label{eq:SaturMMT} \\[0.3em]
\Psinu|_{t=0} &= \Psisig|_{t=0} = \psi_0\in C^\infty(\T),
\end{align}
\end{subequations}
where $\nu>0$ is the viscosity and $\sigma>0$ is the saturation parameter. We denote by $\Psinu$ and $\Psisig$ the solutions of the viscous and saturating problems, respectively. These two classes of regularization are expected to generate qualitatively different singular structures, as discussed in \cite{Fib15} and illustrated in Fig.~\ref{fig:SpaceTimeIllustr}. Additionally, \eqref{eq:ViscousMMT} breaks down the Hamiltonian structure of \eqref{eq:MMT} while \eqref{eq:SaturMMT} remains Hamiltonian but with a Hamiltonian \eqref{eq:HamilSatur} given in Appendix~\ref{sec:NoBlowSatur}. Whenever the discussion applies to both regularizations, we use the compact notation $\Psieps$ where $\varepsilon \in \lbrace \nu,\sigma \rbrace$ and refer to the corresponding generic equation as \epsfNLSref{}. More generally $\varepsilon$ will refer in the following to either regularization mechanism. In all the following, the initial condition is the real, nonnegative smooth bump
\begin{equation}
\label{eq:InitCond} \psi_0(x)=\left(\frac{\mathcal{M}_0}{\M{b}}\right)^{1/2}b(x-\pi), \quad b(x)=
\begin{cases}
\exp\!\left(1-\dfrac{1}{1-(2x)^2}\right),&|2x|<1,\\
0,&|2x|\geq 1,
\end{cases}
\end{equation}
where $\mathcal{M}_0= \M{\psi_0}$ is the mass of the initial condition. This smooth and compactly supported initial condition is symmetric with respect to $\pi$ and the time evolution preserves this symmetry. This single bump initial condition gives a controlled laboratory of the dynamics. The dependence of the limiting behavior on the initial condition is in itself an interesting question that we leave for future work, see for instance the case of KPZ equation leading to several universality subclasses \cite{Cor12,Tak18} The numerical integration for both regularized dynamics uses a Fourier pseudo-spectral method with $N$ equidistant points on $\T$. The fractional operator and spatial derivatives are evaluated spectrally, while the nonlinear terms are evaluated in physical space and projected after each evaluation onto $|k|<N/4$. This half-band truncation removes cubic aliases, and the same projection is applied to the saturating nonlinearity. Both regularizations are advanced with the fourth-order exponential time-differencing Runge-Kutta method (ETDRK4) of Cox and Matthews \cite{CoxMat02}. With our choice $\alpha=1/2$, the linear multipliers are $-i|k|^{1/2}-\nu k^2$ for \eqref{eq:ViscousMMT} and $-i|k|^{1/2}$ for \eqref{eq:SaturMMT}. The ETDRK4 coefficients are evaluated by the contour method of Kassam and Trefethen \cite{KasTre05} to avoid cancellation for small arguments. The vanishing-regularization computations are carried out in double-double precision (about 31 decimal digits), including the Fourier transforms and nonlinear evaluations, with resolutions up to $N=2^{25}$. Such high precision and resolution are necessary to resolve faithfully the collapse event. In particular, the double-double precision was found to be necessary to preserve the symmetry of the solution after regularized collapse events. All computations use $\alpha=1/2$ and $T=2$.
In Figure~\ref{fig:SpaceTimeIllustr}, we display the spatio-temporal evolution of the solution to \epsfNLSref{} for each regularization mechanism and, in each case, two values of the regularization parameter. This figure already contains, at a qualitative level, the different ingredients studied below. The first observation it suggests is the drastic difference between the viscous and saturating regularizations after the first collapse at $t=t^\star\simeq1.012$. Before $t^\star$, the solution of the inviscid problem \eqref{eq:MMT} is unique and both limits converge toward it. After the first blowup, however, the two regularizations lead to qualitatively different behaviors. In the viscous case, the first collapse is sharply localized in space and time and triggers many further collapse events, equally localized. This follows from the dissipative nature of the regularization and ultimately leads to anomalous dissipation. The saturating nonlinearity, by contrast, is conservative, so that the singular structures it produces can be long-lived. Already at this level, the nature of the regularization is seen to matter for the post-blowup life of the solution. For a detailed discussion of continuation under dissipative and conservative regularizations in NLS, see \cite{Fib15}. The second observation is that the number and the location of the singular events depend on the level of regularization. As the inviscid limit is taken, more and more collapse events are triggered after the first one. We will show that this mechanism, referred to below as \emph{collapse proliferation}, is at the root of both anomalous dissipation and spontaneous stochasticity in this system: each singular event dissipates a vanishing amount of mass and generates a vanishing amount of uncertainty in the limit, but their growing number compensates for their vanishingly small individual effect. The spatial organization of secondary collapses may be influenced by dispersive transport. The dispersion relation $\omega(k)=|k|^\alpha$ gives a group velocity $v_g(k)=\alpha\,\operatorname{sgn}(k)|k|^{\alpha-1}$ for $k\neq0$. For $0<\alpha<1$, the smaller the scale of a wave packet, the slower it propagates. As viscosity decreases, collapse is arrested at progressively smaller scales, so that wave packets at these scales have a smaller group velocity. This may contribute to the slower initial expansion of the collapse remnants, and is consistent with the narrower depleted region observed at smaller viscosity in Figure~\ref{fig:SpaceTimeIllustr}. The remnants may subsequently refocus away from the initial collapse centre and trigger secondary collapses. This interpretation remains heuristic: neither the locations of these events nor the specific role of nonlocality are established by the group-velocity argument alone and possible non-local effects are not taken into account.
\subsection{Regularizations prevent blowup}

To fulfill \ref{it:spst-regularized}, the regularized problems must be globally well posed. Proposition~\ref{prop:GWP} in Appendix~A establishes existence, uniqueness, and continuous dependence on the initial data on the torus for every fixed positive regularization parameter. For smooth initial data, the solutions remain smooth for all forward times. The viscous result holds for $\alpha\geq1/2$, including the value used in our simulations, and the saturating result holds for every $\alpha>0$. In particular,
\begin{equation}
\label{eq:BoundHs}
\begin{aligned}
\|\Lambda^{\alpha/2}\Psinu(t)\|_{L^2(\T)}^2 &\leq\|\Lambda^{\alpha/2}\psi_0\|_{L^2(\T)}^2 \exp\!\left(\frac{C_\alpha t}{\nu}\mathcal M_0^2\right),\\
\|\Lambda^{\alpha/2}\Psisig(t)\|_{L^2(\T)}^2 &\leq\|\Lambda^{\alpha/2}\psi_0\|_{L^2(\T)}^2 +\frac{\mathcal M_0}{2\sqrt\sigma}.
\end{aligned}
\end{equation}
Here $\mathcal M_0=\M{\psi_0}$ is the initial mass and $C_\alpha$ depends only on $\alpha$ and the fixed torus. The bound for the viscous regularization is adapted from the argument of \cite{PasSul05} written for two the dimensional cubic NLS. The saturation estimate follows from conservation of its Hamiltonian \eqref{eq:HamilSatur}. The precise global well-posedness statement is given in Proposition~\ref{prop:GWP} together with its proof in the same Appendix~\ref{app:GWP}. The bounds given in \eqref{eq:BoundHs} ensure that blowup is prevented for any finite value of the regularization parameter and therefore ensure that there is a unique solution to the regularized problem \epsfNLSref{} at any finite time. That is the second part \ref{it:spst-regularized} of the Spontaneous Stochasticity triptych. These bounds are finite provided that $\varepsilon>0$, this makes the inviscid limit $\varepsilon \to 0$ a central question, in particular it enables the study of the limiting behavior of the solution beyond the first inviscid blowup time.

Before the inviscid blowup time $t^\star$, there exists a unique strong solution $\psi$ of \eqref{eq:MMT}. It is therefore expected that both regularizations \epsfNLSref{} recover this inviscid solution on any time interval $[0,T]$ with $T<t^\star$, in the sense that
\[
\lim_{\varepsilon \rightarrow 0} \sup_{0\leq t \leq T}\bigl\| \Psieps(t,\cdot)-\psi(t,\cdot)\bigr\|_{L^2_x } =0, \quad \varepsilon \in \lbrace \nu,\sigma \rbrace.
\]
This pre-collapse stability is consistent with weak-strong uniqueness: a weak solution satisfying the required regularity and space-time bounds coincides with the strong solution issued from the same initial condition, for as long as the latter exists.\footnote{For the framework of space-time Strichartz estimates, see \cite{SulSul07} for standard NLS and \cite{HonSir15} for the fractional theory on $\R^d$. The required bounds depend on the dispersion and the spatial domain. Here $d=1$ and $\alpha=1/2$, so the problem is mass-supercritical and energy-critical, and the relevant bounds are space-time norms $L^q_tL^r_x$ with exponents fixed by $\alpha$. } Thus, before blowup, the vanishing-regularization limit is expected to yield the same solution independently of the regularization mechanism. The post-blowup regime $t>t^\star$ is different: the strong-solution comparison no longer applies. Although $H^{\alpha/2}$ is the natural energy space, a post-collapse continuation need not have finite norm in that space. We use $L^2$ for the comparisons both before and after collapse because it remains meaningful for finite-mass fields; its squared distance measures integrated field uncertainty. A Sobolev gap can be dominated by differences in increasingly narrow cores whose mass vanishes, even when the fields approach the same $L^2$ limit. This raises the central question of the present work: Does the regularized dynamics \epsfNLSref{} select a unique inviscid continuation in $L^2$ in the vanishing-regularization limit? If the answer is positive for a given regularization, then that procedure acts as a selection principle. If the regularization fails to select a unique inviscid solution, it might still select a well-defined probability distribution on the space of solutions of the inviscid model, as explained in Section~\ref{sec:SpTrip}. In that scenario, a vanishingly small perturbation of the regularization parameter or of the initial condition can generate an $\mathcal{O}(1)$ difference in the solution in finite time.
\section{Inviscid blowup and anomalous dissipation}
\label{sec:BowupAD}

The first step toward spontaneous stochasticity is the existence of several continuations of the inviscid problem. For the fractional NLS \eqref{eq:MMT}, the post-blowup regime is a natural candidate, since uniqueness of weak solutions may fail. Before the inviscid blowup time $t^\star$, the inviscid limit of \epsfNLSref{} exists and coincides with the unique strong pre-blowup solution of \eqref{eq:MMT}. Our first objective is therefore to provide numerical evidence that finite-time blowup is indeed recovered in the vanishing-regularization limit for both models.
\subsection{Finite time blowup in the inviscid limit}

To this end, we monitor the contribution of the linear part of the dynamics to the Hamiltonian \eqref{eq:Hamil} and the supremum norm of the solution,
\begin{equation}
\label{eq:Halphanorm} \Sobol{\Psieps}(t):= \int_{\T} \bigl| \Lambda^{ \frac{\alpha}{2}}\Psieps(t,x) \bigr|^2 \, dx =2\pi \sum_{k\in\mathbb{Z}} |k|^{\alpha}\,|\widehat{\psi}_\varepsilon(t,k) |^2, \qquad \|\Psieps(t)\|_\infty = \sup_{x \in \T} |\Psieps(t,x)|,
\end{equation}
and study their behavior as $\varepsilon\to 0$. We will also refer to $\mathcal H_L$ as the critical Sobolev norm. These quantities diagnose the concentration described in Section~\ref{sec:InviscidModel}; divergence of the energy norm is not required for collapse at energy criticality.

\begin{figure}[htb!]
\centering
\includegraphics[width=0.95\linewidth]{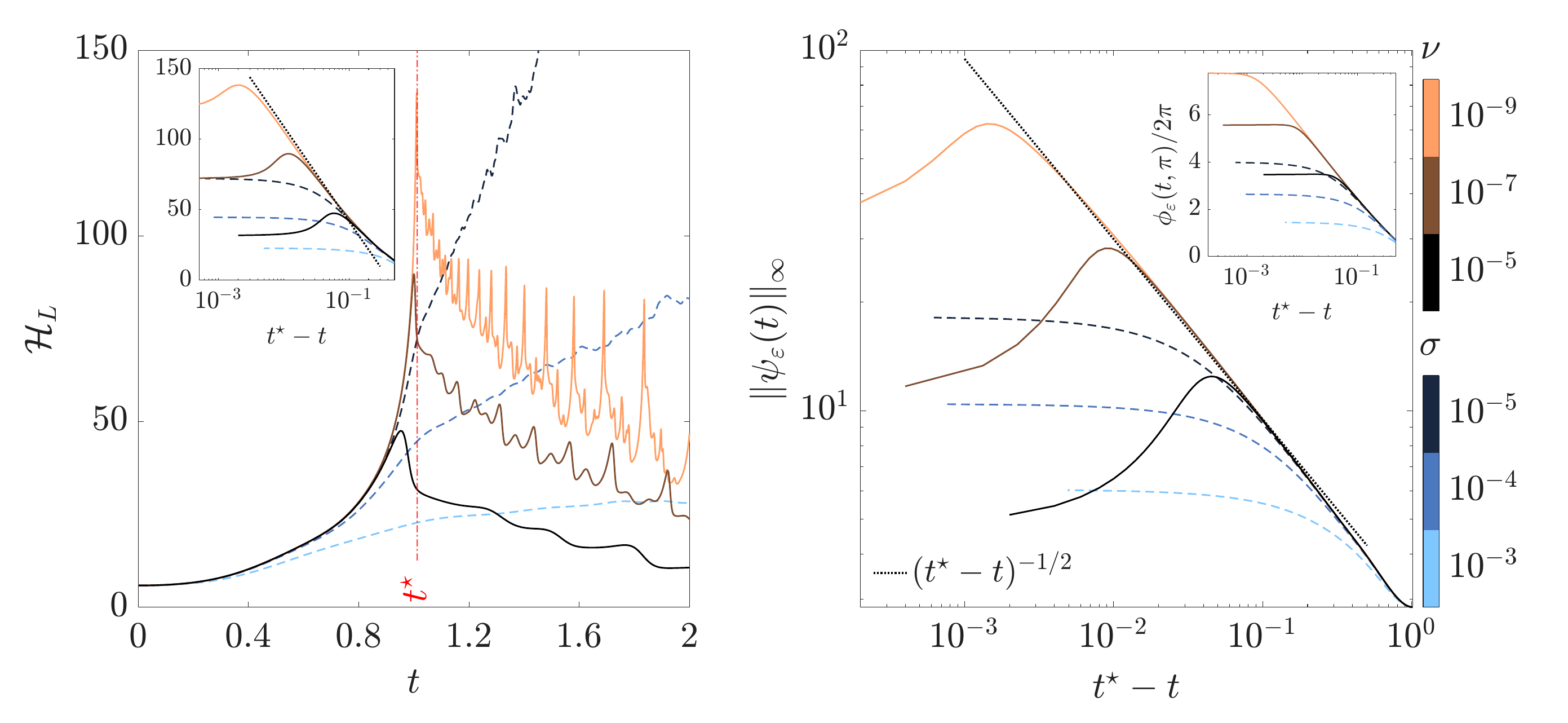}
\includegraphics[width=0.95\linewidth]{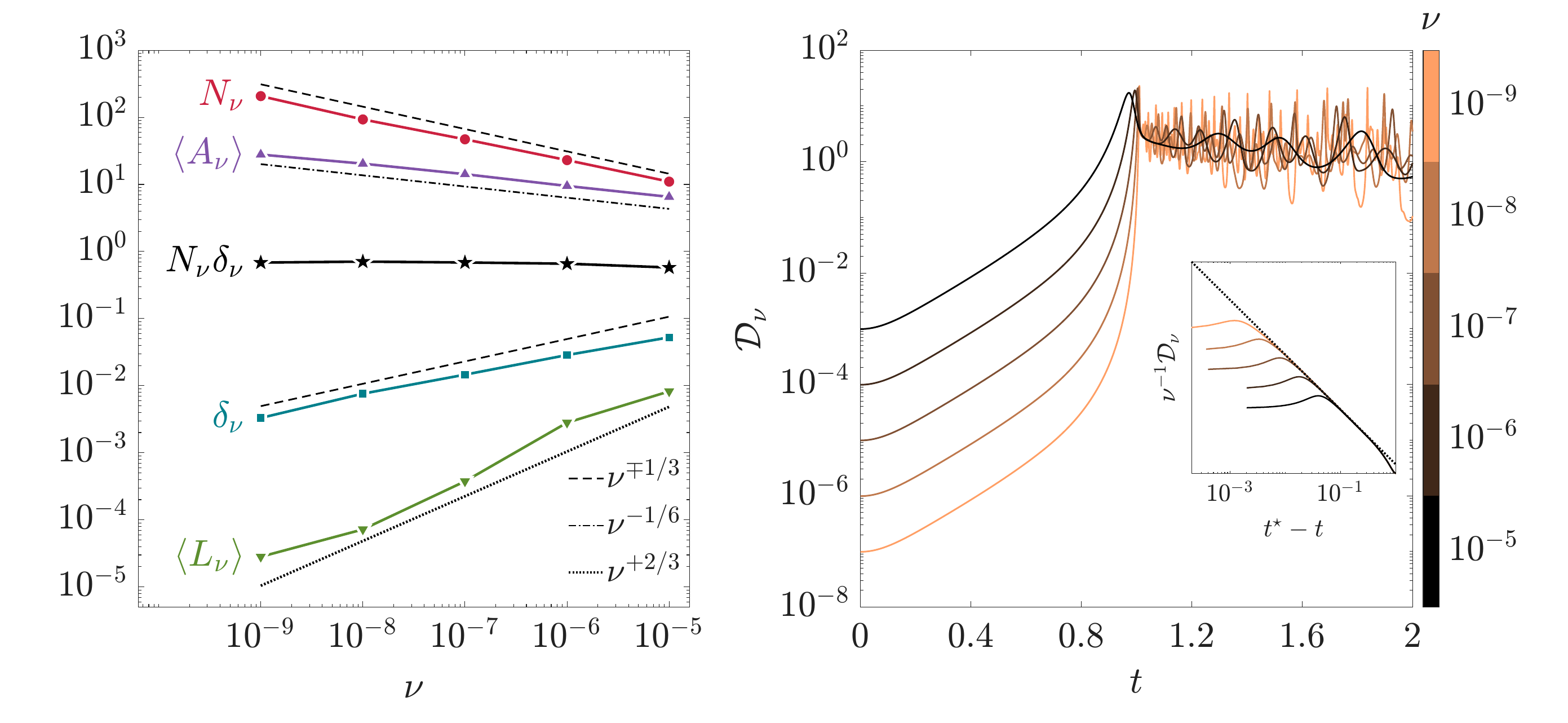}
\caption{Top left: linear part of the Hamiltonian $\mathcal{H}_L$ \eqref{eq:Halphanorm} as a function of time for decreasing $\varepsilon$. Dashed blue curves correspond to the saturating model \eqref{eq:SaturMMT}, and solid black-to-orange curves to the viscous model \eqref{eq:ViscousMMT}. The colorbars give the three values of each regularization parameter; the red vertical line marks $t^\star$. As $\varepsilon\to 0$, all curves collapse onto the same pre-blowup behavior up to $t^\star\simeq 1.012$, after which $\mathcal{H}_L$ diverges in the inviscid limit. The inset shows $\mathcal{H}_L$ against $t^\star-t$ on a semi-logarithmic scale, advocating in favor of a logarithmic divergence of the critical norm. Top right: maximum amplitude of the solution as a function of $t^\star-t$ for both regularizations. The colors match those of the left plot. The black dotted line shows the scaling $(t^\star-t)^{-1/2}$. The inset shows the phase accumulated at the collapse point, $\phi_\varepsilon(t,\pi)/2\pi$, as a function of $t^\star-t$ showing a logarithmic divergence of the accumulated phase upon collapse. Bottom left: mechanism of the proliferation-induced anomalous mass dissipation, as a function of the viscosity. Red circles: total number of collapse cores. Teal squares: heuristic estimate of the mass dissipated per core. Black stars: total mass dissipated in the cores. The purple (resp. green) triangles corresponds to the average maximum amplitude (resp. minimum core size) attained by collapsing cores. The black dashed, dash-dotted and dotted lines show the expected scalings $\nu^{\mp1/3}$, $\nu^{-1/6}$ and $\nu^{2/3}$, respectively. Bottom right: mass dissipation rate $\mathcal{D}_\nu(t)$ in the viscous case. Before blowup, the dissipation rate vanishes proportionally to $\nu \to 0$ (at fixed $t < t^\star$). After blowup, it reaches a finite value independent of viscosity; this is anomalous dissipation. The inset displays the pre-blowup behavior of $\nu^{-1}\mathcal{D}_\nu$ as a function of $t^\star-t$ for vanishingly small values of the viscosity. The dotted black line shows the $-3$ power law scaling expected from heuristic arguments presented in the main text. For an illustration of inviscid wave collapse, see \href{https://youtu.be/27GHylPq4BE}{this video}.}
\label{fig:BlowupAnomDiss}
\end{figure}

To characterize the approach to blowup, we use the collapse scalings conjectured in the numerical study \cite{KleSpa14} of the fractional NLS \eqref{eq:MMT}. For a single-bump initial condition in the regime $\alpha<1$ considered here, the peak amplitude and the typical core width $L(t)$ are expected to satisfy
\begin{equation}
\label{eq:ConjBlowupPsi} \| \psi(t)\|_\infty \underset{t \to t^\star}{\propto} (t^\star-t)^{-\frac{1}{2}},~~~~ L(t) \underset{t \to t^\star}{\propto} (t^\star-t)^{\frac{1}\alpha}.
\end{equation}
The upper row of Fig.~\ref{fig:BlowupAnomDiss} gathers numerical results showing that in the inviscid limit of \epsfNLSref{}, wave collapse is restored. We compute the norms \eqref{eq:Halphanorm} for vanishingly small values of the regularization parameters given in the colorbars to the right of the upper row. The upper-left panel shows the evolution of the critical norm $\mathcal{H}_L$ for the viscous regularization (solid black-to-orange curves) and the saturating regularization (dashed blue curves). The upper-right panel displays $\|\psi_\varepsilon\|_\infty$ as a function of $t^\star-t$ and compares its behavior to the expected inverse square root \eqref{eq:ConjBlowupPsi} scaling shown by the black dotted line.
For early times $t<t^\star$, the curves corresponding to different values of $\nu$ and $\sigma$ collapse onto a single profile in the limit, independently of the regularization mechanism. This is consistent with convergence toward the same pre-blowup strong solution of the inviscid equation \eqref{eq:MMT}. Beyond $t^\star$, the behavior becomes regularization-dependent: $\mathcal{H}_L$ remains bounded for each fixed $\nu$ or $\sigma$, but grows without apparent bound as the regularization parameters decrease. This is consistent with the bounds \eqref{eq:BoundHs}, which allow $\Sobol{\Psieps}$ to grow like $e^{ct/\nu}$ or $c\sigma^{-1/2}$, and with the occurrence of blowup in the inviscid limit. A more detailed analysis, not shown here, reveals that in our simulations the maximum of $\mathcal{H}_L$ over time grows more slowly than the worst-case estimates $e^{ct/\nu}$ and $\sigma^{-1/2}$ for \eqref{eq:ViscousMMT} and \eqref{eq:SaturMMT}, respectively. The inset resolves the approach to $t^\star$ on a logarithmic scale. As discussed in the heuristic of Section~\ref{sec:InviscidModel}, at energy criticality $\alpha=1/2$, the critical norm diverges slower than any power law. The curves of the inset of the upper left panel of Figure~\ref{fig:BlowupAnomDiss} display a logarithmic divergence until each regularization arrests the collapse, which is consistent with this heuristic.

The upper-right panel of Figure~\ref{fig:BlowupAnomDiss} complements this picture by showing the maximum amplitude $\|\Psieps\|_\infty$ as a function of $t^\star-t$. As mentioned above, the expected amplitude growth rate for a single bump initial condition from \cite{KleSpa14} is $(t^\star-t)^{-1/2}$. The superimposed power law shows good agreement with our numerical results. As we will see on several occasions throughout the present paper, this blowup rate can be used at a heuristic level to make several predictions on the pre-blowup behavior of the solution. The blowup time is estimated using the growth rate of the amplitude of the solution with viscous regularization at the smallest accessible viscosity. Postulating the inverse-square-root behavior \eqref{eq:ConjBlowupPsi}, we then fit the blowup time. This procedure is validated a posteriori since the amplitude follows the scaling \eqref{eq:ConjBlowupPsi} with our estimated $t^\star\simeq 1.012$. A more involved approach to estimate this blowup time more precisely would be to use complex plane singularities \cite{PikBar24}. We are however not pursuing an exhaustive characterization of the inviscid blowup in the present work and leave this for future investigations. The inset of the same panel shows the phase accumulated at the collapse point, $\phi_\varepsilon(t,\pi)=\arg\psi_\varepsilon(t,\pi)$. It grows logarithmically as $t\to t^\star$ and $\varepsilon\to 0$. This is the loss of phase reported for the standard NLS \cite{FibKle11,ShiSch12}; a related loss of predictability arises when the same dynamics is integrated backwards \cite{SagDitGoo20}. Altogether, these results are consistent with the restoration of blowup in the inviscid limit for each regularization mechanisms. Both the amplitude of the solution and the mass of its derivative of order $\alpha/2$ diverge in finite time in the inviscid limit. At $t^\star$, the solution of \eqref{eq:MMT} with initial condition $\psi_0$ loses its initial regularity. Past this time, weak solutions come into play: one requires only finite mass, not differentiability. This flexibility in the notion of solution allows nonuniqueness past blowup, as explored in the following sections.
\subsection{Anomalous dissipation in the inviscid limit}

In addition to the possible breakdown of uniqueness, any property of the solution that relies on regularity may cease to be verified after blowup. In particular, mass need not be conserved. In order to assess whether mass is conserved in the inviscid limit or not, we compute both the total mass, for both regularizations, and the mass dissipation rate for the viscous regularization
\begin{equation}
\label{eq:DefDissipRate} \mathcal{D}_\nu(t)= 2\nu \int_\T \bigl|\partial_x \Psinu(t,x)\bigr|^2 \, dx.
\end{equation}
Multiplying \epsfNLSref{} by the complex conjugate of the solution leads to the mass budgets
\begin{equation}
\label{eq:MassBudg}
\begin{array}{ccc}
\displaystyle\frac{\mathrm{d}}{\mathrm{d}t}\M{ \Psinu(t)}= -\mathcal{D}_\nu(t), &~~~~&\displaystyle\frac{\mathrm{d}}{\mathrm{d}t}\M{ \Psisig(t)}=0.
\end{array}
\end{equation}
Thus, for fixed $\nu>0$, viscosity can only dissipate mass, whereas the saturating nonlinearity conserves it. Before blowup, the inviscid solution is smooth, mass conservation is therefore recovered in the vanishing-viscosity limit for $t<t^\star$. After blowup, however, there is no such guarantee. By contrast, the conservative regularization \eqref{eq:SaturMMT} preserves mass for all $\sigma>0$, and its vanishing-$\sigma$ limit therefore conserves the initial mass. The bottom row of Figure~\ref{fig:BlowupAnomDiss} displays numerical results concerning mass conservation. As sketched in the heuristic argument of Section~\ref{sec:InviscidModel}, for $\alpha <1$, a collapsing core carries a mass$\propto L^{1-\alpha}\to 0$ when $t \to t^\star$. A single collapsing core can therefore a priori not dissipate mass in the limit. How then can mass dissipation be sustained in the limit? The bottom left panel of Figure~\ref{fig:BlowupAnomDiss} answers this: each collapse event dissipate a vanishing amount of mass but the number of collapse event diverges in the inviscid limit. Let us sketch here the effect of viscosity on the collapse, leading to an estimation of the total mass dissipated by each collapse for a given viscosity. First of all, the collapse is arrested at a scale $L_\nu$ where dissipative and dispersive effects are comparable, namely $ \nu L_\nu^{-2} \sim L_\nu^{-\alpha}$. Once the core reaches the size $L_\nu$ it will dissipate mass for the amount of time $\tau_\nu$ needed for the inviscid core to shrink from $L_\nu$ to $0$. Namely $\tau_\nu \sim L_\nu^\alpha$. Finally, the amplitude reached by the core of size $L_\nu$ is $A_\nu \sim L_\nu^{-\frac{\alpha }{2}} $. The total amount of mass $\delta_\nu$ dissipated during $\tau_\nu$ will therefore scale as $\delta_\nu \sim \nu \sfrac{A_\nu^2}{L_\nu} \times \tau_\nu $. Overall, we obtain,
\begin{equation}
\label{eq:ViscousScaling} L_\nu \sim \nu^{\frac{1}{2-\alpha }}, \quad \tau_\nu \sim \nu^{\frac{\alpha}{2-\alpha }}, \quad A_\nu \sim \nu^{-\frac{\alpha}{2(2-\alpha) }}, \quad \delta_\nu \sim \nu^{\frac{1-\alpha}{2-\alpha}}.
\end{equation}
Therefore, for $\alpha =1/2$, the number of collapse event $N_\nu$ needed to sustain mass dissipation in the inviscid limit must grow as $\delta_\nu^{-1}\sim \nu^{- \frac{1}{3}}$. The bottom left panel of Figure~\ref{fig:BlowupAnomDiss} precisely support this scaling. We display with a teal line and square markers the average dissipated mass per collapse event as a function of viscosity and compare it favorably to the expected $\nu^{1/3}$ from weak collapse heuristics \eqref{eq:ViscousScaling}. At the same time, the total number of collapse event $N_\nu$ grows as $\nu^{-\frac13}$ as shown by the red line with circle markers, that is the quantitative counterpart of the qualitative Figure~\ref{fig:SpaceTimeIllustr} showing that a multitude of collapsing cores are generated after the inviscid blowup time $t^\star$ and that their number increases in the inviscid limit. The rough estimation of total mass dissipation given by $N_\nu \times \delta_\nu $ represented in black line with star markers remains visibly constant when the viscosity vanishes. That is supported by the exact computation of mass drop (not shown) which is the time integrated version of the bottom left panel of Figure~\ref{fig:BlowupAnomDiss}, showing a sustained mass drop in the limit as a consequence of a sustained dissipation rate shown in this panel. We also display in the bottom left panel of Figure~\ref{fig:BlowupAnomDiss} the scaling with viscosity of the average peak amplitude $\left \langle A_\nu \right \rangle$ of collapsing cores (purple upward triangles) and the average size of cores at peak amplitude $\left \langle L_\nu \right \rangle$ (green downward triangles). They both show a good agreement with the heuristic scalings \eqref{eq:ViscousScaling}. More precise than the crude estimate we just presented, the bottom right panel displays the mass dissipation rate $\mathcal{D}_\nu$ as function of time for all simulated viscosities. Although $\mathcal{D}_\nu(0)=2\nu\|\partial_x\psi_0\|^2_{L^2}$ vanishes as $\nu\to 0$, the dissipation rate remains finite after blowup. A nonzero amount of mass is therefore irreversibly lost in the inviscid limit. Similar phenomena have been observed in focusing NLS ($\alpha=2$) \cite{JosPom20,AmaJos23,PasSul05}; we are not aware of such results in the fractional, unforced, case considered here. This difference between viscous and saturating regularizations explains the qualitatively different behaviors of the solution past blowup in Figure~\ref{fig:SpaceTimeIllustr}. In the viscous case, focusing of mass leads to locally extreme events of mass dissipation. In that regard, the singularities are short-lived under viscous regularization, which explains the space-time Dirac-like structure of the second panel of Figure~\ref{fig:SpaceTimeIllustr}. On the other hand, the saturating nonlinearity prevents blowup not by dissipating energy, but by hindering the underlying instability. This leads to long-lived singular structures as observed in Figure~\ref{fig:SpaceTimeIllustr} and justifies that post blowup continuations will strongly depend on the regularization mechanism. Thanks to the expected blowup behavior \eqref{eq:ConjBlowupPsi} of the inviscid solution, one expects the following behavior for the mass dissipation rate
$$
\lim_{\nu \to 0} \nu^{-1} \mathcal{D}_\nu(t)\underset{t \to t^\star}{\propto} \left(\frac{\|\psi\|_\infty}{L(t)}\right)^2L(t)\underset{t \to t^\star}{\propto} (t^\star -t)^{- \frac 1\alpha-1}.
$$
With our choice $\alpha=1/2$, this leads to a $ \nu (t^\star -t)^{-3}$ behavior of the pre-blowup dissipation rate, which is indeed observed in the inset of the bottom right panel of Figure~\ref{fig:BlowupAnomDiss} where we display the rescaled dissipation rate $\nu^{-1}\mathcal{D}_\nu$ as a function of the distance to blowup time and compare it to the $-3$ power law scaling.
\subsection{Mass defects and flexibility of the post-blowup state}
\label{sec:FlexPostBlowup}

The previous subsection shows that the viscous approximation retains a finite mass dissipation rate in the inviscid limit. This observation raises a structural question. The inviscid equation \eqref{eq:MMT} formally conserves mass, so how can an inviscid limiting continuation lose mass? The point is that the formal conservation law is justified only as long as the solution is regular enough. After collapse, this regularity is lost, and mass conservation becomes a property of the selected weak continuation rather than a direct consequence of the inviscid equation alone. We make this mechanism explicit through a coarse-grained formulation, in the spirit of the Duchon-Robert and Constantin-E-Titi approaches to anomalous dissipation in fluid turbulence \cite{DucRob00,ConE94,Eyi94}. Let \(\theta\) be a smooth periodic mollifier with unit integral, set \(\theta_\ell(x)=\ell^{-1}\theta(x/\ell)\), and write \(\psi_\ell=\theta_\ell\ast\psi\). If \(\psi\) solves the inviscid equation \eqref{eq:MMT}, then the filtered field is smooth and satisfies the non-closed dynamics \(i\partial_t\psi_\ell=\Lambda^\alpha\psi_\ell-\left(|\psi|^2\psi\right)_\ell\). Taking the time derivative of the mass of the coarse-grained field yields
\begin{align}
\label{eq:MassFilteredRaw} &\frac{d}{dt}\M{\psi_\ell} = -2\,\Im\int_\T \overline{\psi_\ell(t,x)} \left[ \left(|\psi|^2\psi\right)_\ell(t,x) - |\psi_\ell(t,x)|^2\psi_\ell(t,x) \right]\,dx = -\mathsf D^\ell[\psi],\\
&\M{\psi_\ell}(t)=\M{\psi_\ell}(0) -\int_0^t \mathsf D^\ell[\psi](s)\,ds \label{eq:MassFilteredRawTimeInteg}
\end{align}
The quantity \(\mathsf D^\ell[\psi]\) measures the nonlinear transfer of mass from scales larger than \(\ell\) to scales smaller than \(\ell\). If \(\mathsf D:=\lim_{\ell\to0}\mathsf D^\ell\) exists and is nonzero, the weak inviscid continuation carries a mass defect. This defect is the analogue of the anomalous energy flux in weak Euler solutions. Such a defect cannot occur while the inviscid solution satisfies $\|\psi(t)\|_{L^4}^4=\int_\T |\psi(t,x)|^4dx <+\infty$ \footnote{ Indeed, at a fixed time, if \(\psi\in L^4(\T)\), then \(\psi_\ell\to\psi\) in \(L^4\), while \(u\mapsto |u|^2u\) is continuous from \(L^4(\T)\) to \(L^{4/3}(\T)\). Hence \(\left(|\psi|^2\psi\right)_\ell-|\psi_\ell|^2\psi_\ell\to0\) in \(L^{4/3}\), and H{\"o}lder's inequality gives \(\mathsf D^\ell[\psi]\to0\)}. Wave collapse provides precisely a mechanism for which such control on the inviscid solution breaks down. Along smooth pre-blowup solutions, the Hamiltonian identity \eqref{eq:Hamil} gives \(\|\Lambda^{\alpha/2}\psi(t)\|_{L^2}^2=2\Ha{\psi_0}+\frac{1}{2}\int_\T |\psi(t,x)|^4dx\). Thus the divergence of \(\|\Lambda^{\alpha/2}\psi(t)\|_{L^2}\) at collapse is accompanied by the divergence of \(\int_\T |\psi(t,x)|^4dx\). Collapse is therefore a natural event at which the nonlinear commutator in \eqref{eq:MassFilteredRaw} may acquire a nonzero inviscid limit. Importantly however, mass can still be conserved even if \(\int_\T |\psi(t,x)|^4\) diverges at the collapse. In particular, if \(\int_0^{t^\star}\int_\T |\psi(s,x)|^4dxds<\infty\), then the time integrated mass defect \eqref{eq:MassFilteredRawTimeInteg} vanishes and mass is conserved upon collapse\footnote{Indeed, let
\[
\mathcal C^\ell[\psi]:=\left(|\psi|^2\psi\right)_\ell-|\psi_\ell|^2\psi_\ell.
\]
If \(\psi\in L^4((0,t^\star)\times\T)\), then \(u\mapsto |u|^2u\) from \(L^4\) to \(L^{4/3}\) gives \(\mathcal C^\ell[\psi]\to0\) in \(L^{4/3}((0,t^\star)\times\T)\). Consequently,
\[
\int_0^{t^\star}\left|\mathsf D^\ell[\psi](t)\right|\,dt\leq2\|\psi_\ell\|_{L^4((0,t^\star)\times\T)}\|\mathcal C^\ell[\psi]\|_{L^{4/3}((0,t^\star)\times\T)}\longrightarrow0.
\]
In particular, a logarithmic divergence of \(\|\psi(t)\|_{L^4(\T)}^4\) remains integrable in time and cannot produce a finite mass jump at \(t^\star\).}. That is precisely the case of weak collapse we are studying here. The individual collapsing cores are not dissipating mass in the limit while the linear part of the Hamiltonian (hence $\int |\psi|^4$) diverges logarithmically at collapse. The logarithmic singularity being integrable in time, no mass is dissipated by a single collapse. The logarithmic divergence is specific to our borderline, energy-critical system ($\alpha =1/2$). For the $d$-dimensional model with nonlinearity $|\psi|^{2p}\psi$ discussed in Section~\ref{sec:InviscidModel}, the same dimensional argument predicts a core contribution to $\|\psi(t)\|_{L^{2p+2}}^{2p+2}$ scaling as $(t^\star-t)^{\frac{d}{\alpha}-\frac{p+1}{p}}$. This power law is integrable in time when $\alpha<pd$, precisely the mass-supercritical regime in which weak collapse is expected. This supports the expectation that an isolated weak collapse produces no finite mass jump, but remains a heuristic extension of the cubic commutator argument above.

These mass-defect considerations also clarify the role of regularization. Before collapse, the solution is smooth and the integrated commutator defect $\lim_{\ell \to 0}\int_0^t \mathsf{D}_\ell(s)ds$ vanishes. In the present energy-critical case, the observed logarithmic divergence of $\mathcal H_L$ is integrable in time and does not by itself produce a finite integrated mass defect at the first collapse. After collapse, the inviscid equation may admit infinitely many weak continuations, and the value of the defect depends a priori on the considered continuation. Regularizing the inviscid equation therefore acts as a way to select qualitative behavior of the post blowup life of the solution. This is exactly what the two collapse-arresting mechanisms display, no single collapse event is sufficient to dissipate mass in the limit but their proliferation compensate the vanishingly small amount of mass each of them dissipates. The inviscid limit of viscous regularization will lie in dissipative continuations while the saturating nonlinearity will remain conservative. This flexibility of the weak post-blowup dynamics is in fact the mirror of the absence of uniqueness of solution of \eqref{eq:MMT} after blowup: two solutions with different masses cannot coincide.
\subsection{Discussion: connections with fluid turbulence}

One reason for choosing \eqref{eq:MMT} as an inviscid model is the extensive literature on finite-time blowup for NLS equations. For fluid models such as the incompressible Euler equations, finite-time blowup results are much scarcer; see however \cite{LuoHou14,CamMai18,AddRic26}. For focusing NLS, the blowup mechanism is well known and the literature on the subject is vast \cite{SulSul07,Fib15}, see also \cite{BanLuc24}. Besides being a potential experimental testbed for spontaneous stochasticity, the present model is more analytically tractable than the full incompressible Euler equations, and may therefore also serve as a useful toy model for spontaneous stochasticity in fluids. The parallel between the two settings is not, however, a straightforward one. In a bounded domain for instance the vanishing-viscosity limit of Navier-Stokes need not converge to a smooth Euler solution, and weak-strong uniqueness is recovered only under additional conditions on the solution at or near the wall \cite{Kat84,BarTit13,Kel17}. Boundary-generated vortex sheets then provide a route to anomalous dissipation and to nonunique limits that requires no finite-time blowup of a smooth Euler solution \cite{BarSzeWie14,VasYan23}. In that setting the dissipation is carried by vorticity generated at the wall and fluxed into the bulk flow, a balance made quantitative by the Josephson-Anderson relation and shown to hold for weak Euler solutions in the zero-viscosity limit \cite{Eyi21,QuaEyi24,Eyi24}.

In fluid turbulence, inviscid energy dissipation is known as anomalous dissipation. It is a direct manifestation of the lack of regularity of turbulent Navier-Stokes solutions and lies at the core of turbulence phenomenology \cite{Fri95,Pop00,RufChe26}, and the results presented here can be seen as its analogue in a dispersive setting \cite{Fri95,Pop00}. In Navier-Stokes turbulence, kinetic energy undergoes anomalous dissipation while the total energy is conserved, the dissipated kinetic energy reappearing as a heat source in the internal-energy balance of the Navier-Stokes-Fourier system \cite{LanLif87,FeiNov17}. In physical systems governed by NLS-type equations, the total mass must likewise be conserved, so that the anomalously dissipated mass has to feed some other physical quantity. In nonlinear optics, mass is proportional to the beam power, which is lost to multiphoton absorption and plasma generation \cite{CouMys07}. In a superfluid, where the term originates and refers to the total mass of the condensate, the dissipated mass is transferred to the normal component, both in the damped Gross-Pitaevskii description and in coupled models that conserve the total mass \cite{Pit59,PitStr16,JayTri22}.

The analogy of the inviscid dissipation mechanisms discussed in \ref{sec:FlexPostBlowup} with Onsager theory of inviscid energy dissipation in fluids \cite{Ons49,Eyi24} is direct. For incompressible Euler, the formal conservation of kinetic energy is valid only above a critical regularity threshold of H{\"o}lder $1/3$ conjectured by Onsager. In the coarse-grained balance, the possible failure of conservation is encoded by a nonlinear commutator akin to \eqref{eq:MassFilteredRaw}. Eyink \cite{Eyi94} and Constantin, E, and Titi \cite{ConE94} proved that this flux vanishes for sufficiently regular weak solutions, while convex-integration constructions show that below Onsager regularity weak Euler solutions may fail to conserve kinetic energy \cite{Ons49,Ise18,BucDeL18}. Later results revealed the flexibility of weak solutions further still, realizing broad classes of energy profiles \cite{BucDeL18,BucVic19}, including increasing ones. In convex-integration constructions, weak Euler or Navier-Stokes solutions are obtained through approximate Euler-Reynolds systems with a stress \(R_q\to0\). That stress is not a physical viscosity but an auxiliary approximation mechanism, and the flexibility of the limiting energy profile reflects the freedom that the low regularity of the solution leaves in the choice of the sequence $R_q$. Apparently unphysical energy behaviors should therefore be read as consequences of selection by an Euler-Reynolds type regularization of the inviscid dynamics, rather than of the inviscid dynamics itself.

The results of this section provide a clear manifestation of post-blowup nonuniqueness for \eqref{eq:MMT}: different regularization mechanisms yield distinct inviscid limits, one conservative and one dissipative. They are therefore necessarily distinct solutions. In the next section, we go one step further and show that neither of these two regularizations acts as a deterministic selection principle justifying the statistical interpretation of the inviscid limits presented in Section~\ref{sec:SpTrip}.
\section{ Breakdown of determinism in the singular limit}
\label{sec:LSP}

The previous section established two key properties of the vanishing-regularization limit. First, both regularizations remain compatible with finite-time blowup of the inviscid dynamics. Second, for the viscous regularization, the inviscid limit exhibits a finite mass dissipation rate, namely anomalous dissipation of mass. In particular, the inviscid limits associated with viscous diffusion and nonlinear saturation cannot coincide after blowup since one dissipates mass, while the other conserves it. These results already show that post-blowup continuations of \eqref{eq:MMT} depend on the regularization mechanism. Such a dependence is not by itself spontaneous stochasticity. For the ideal Burgers equation, vanishing viscosity selects the unique entropy solution, whereas the zero-dispersion limit of KdV selects a distinct, measure-valued weak solution \cite{LaxLev79,Ven85}: the two limits differ, yet each regularization mechanism select a given solution; see nonetheless \cite{EyiDri15} for the Lagrangian spontaneous stochasticity of the backward particle trajectories of the same equation, which coexists with this deterministic Eulerian limit. Whether a fixed regularization selects a unique continuation must therefore be tested separately, which is the object of this section. We now ask whether a post-blowup continuation is selected once the regularization mechanism has been fixed. This is the first deterministic test suggested by the spontaneous-stochasticity framework of Section~\ref{sec:SpTrip}. If \(\psi_\varepsilon(t)\) denotes the solution of a well-posed regularized dynamics, meaning that the solution is unique and depends continuously on the initial condition, the inviscid limit might thus fail in two ways. First, uniqueness might be lost in the limit, meaning that \(\psi_\varepsilon(t)\) has no limit when $\varepsilon \to 0$. Second, a perturbation of the initial condition vanishing with $\varepsilon$ might still produce an $\mathcal{O}(\varepsilon^0)$ effect on the solution at finite time. Note that these two aspects are a priori non-equivalent and we will explore both in the following. In the measure-theoretic formulation developed in \cite{RufSim26} and recalled in Section~\ref{sec:SpTrip}, it is precisely this breakdown of a deterministic limiting procedure that gives rise to the notion of Spontaneous Stochasticity. In this section, we will provide numerical evidence that for each regularization mechanism, the inviscid limit fails in both regards. This implies that the inviscid limit is better described as a random process within the spontaneous stochasticity framework, making this result the central one of this paper.
\subsection{Selection of an inviscid solution}
\label{sec:LSP_reg}

We first test whether the regularized dynamics selects a unique inviscid continuation when the initial condition is kept fixed. To this end, we compare two solutions of \epsfNLSref{} with different regularization parameters (but the same regularization mechanism). For $(\varepsilon_1,\varepsilon_2)= (\nu_1,\nu_2)$ or $(\varepsilon_1,\varepsilon_2)= (\sigma_1,\sigma_2)$ we define the mass gap measuring the mass of the discrepancy between two solutions,
\begin{equation}
\label{eq:MassErrorParamReg} \mathcal{G}_{\varepsilon_1,\varepsilon_2}(t):=\M{\psi_{\varepsilon_1}(t)-\psi_{\varepsilon_2}(t)}.
\end{equation}
Before the first blowup time $t^\star$, the inviscid solution of \eqref{eq:MMT} is unique. Therefore one can show (not detailed) that
\[
\lim_{\varepsilon_1,\varepsilon_2\to 0}\mathcal{G}_{\varepsilon_1,\varepsilon_2}(t)=0, \qquad t<t^\star.
\]
The post-blowup regime is qualitatively different. If, for some $t>t^\star$, the limit of $\mathcal{G}_{\varepsilon_1,\varepsilon_2}(t)$ remains strictly positive \footnote{Note that the limit does not necessarily exist; one can just replace it by $\liminf_{\varepsilon_1,\varepsilon_2\to 0}\mathcal{G}_{\varepsilon_1,\varepsilon_2}$} as $\varepsilon_1,\varepsilon_2\to 0,\; \varepsilon_1 \neq\varepsilon_2 $, then the corresponding family of regularized solutions does not converge in $L^2$ toward a unique solution of the inviscid problem \eqref{eq:MMT}. In that case, the regularization fails to act as a selection principle. Although it produces a unique solution for every fixed $\nu>0$ or $\sigma>0$, it does not select a unique inviscid continuation when the regularization is vanishing. That is the breakdown of the deterministic interpretation of the inviscid limit as sketched in Section~\ref{sec:SpTrip}. In other words, the lack of selection principle amounts to the solution remaining drastically sensitive to the precise value of the regularization parameter, even in the limit. The regularization level is not the only possible source of extreme sensitivity, one can also explore the stability with respect to perturbations of the initial condition.
\subsection{Sensitivity to the initial condition}
\label{sec:LCWRTIC}

The sensitivity of the final state to an initial perturbation is of central importance when trying to predict the evolution of a system. In his seminal work, Lorenz \cite{Lor69} understood that multiscale dynamics could lead to an amplification of an initial error such that the overall error on the final state could not be reduced by reducing the error on the initial state, a property stronger than deterministic chaos, also famously introduced by Lorenz. Such infinite sensitivity corresponds to a lack of continuity with respect to the initial condition. Most studies on spontaneous stochasticity exploit this lack of continuity with respect to the initial condition. See for instance the case of two dimensional Kelvin-Helmholtz instability \cite{ThaBec20} which undergoes the so called norm-inflation phenomenon \cite{Leb02} also present in dispersive settings \cite{RufTzv26}. To study the sensitivity of the inviscid limit to the approximation of the initial condition, we use a regularization-dependent randomization of the initial datum,
\begin{equation}
\label{eq:RandomInit} \psi_0^{\varepsilon,\omega}(x) = \sum_{k\in\mathbb{Z}} e^{i\varepsilon^{\lambda}\theta_\omega(k)}\,\widehat{\psi}_0(k)e^{i kx},\quad \varepsilon\in\lbrace \nu,\sigma \rbrace
\end{equation}
where the $\theta_\omega(k)$ are i.i.d. random variables uniformly distributed on $[0,2\pi]$ and $\lambda >0$ is a free exponent to be chosen. This choice has two useful properties. First, it preserves the mass exactly for any $\varepsilon$ and any realization $\omega$, $ \M{\psi_0^{\varepsilon,\omega}}=\M{\psi_0}. $ Second, the perturbation is controlled as $\varepsilon\to 0$. Indeed, for every $x\in\T$ and every realization $\omega$,
\begin{equation}
\label{eq:BoundRandomInit} |\psi_0^{\varepsilon,\omega}(x)-\psi_0(x)| \leq 2\pi \varepsilon^{\lambda} \sum_{k\in\mathbb{Z}} |\widehat{\psi}_0(k)|.
\end{equation}
Therefore the random initial condition converges pointwise to $\psi_0$ as the regularization vanishes, for every realization of the random phases as long as the Fourier modes of the initial condition are summable, which is the case for a smooth initial condition. To measure sensitivity to this vanishing perturbation, we consider two statistically independent realizations $\Psieps^{\omega_1}$ and $\Psieps^{\omega_2}$ of either regularized dynamics, with the same regularization parameter $\varepsilon$ but independent randomized initial conditions. We define the associated mass gap
\begin{equation}
\label{eq:MassErrorRandomInit} \widetilde{\mathcal{G}}_{\varepsilon}(t,\omega_1,\omega_2) := \M{\Psieps^{\omega_1}(t)-\Psieps^{\omega_2}(t)},\quad \varepsilon\in\lbrace \nu,\sigma \rbrace.
\end{equation}
Two initial conditions that become indistinguishable as $\varepsilon\to 0$ may however generate macroscopically different post-blowup states. At this stage, $\widetilde{\mathcal{G}}_{\varepsilon}(t)$ is a random variable and we do not yet average it. The corresponding statistical description is discussed in the next section.
\begin{remark*}
Sensitivity to the regularization parameter and sensitivity to the initial condition are distinct phenomena. The passive scalar example of Armstrong and Vicol \cite{ArmVic25} illustrates this point. It obeys $\partial_t\theta^\kappa+b\cdot\nabla\theta^\kappa=\kappa\Delta\theta^\kappa$, where $b$ is an incompressible multiscale vector field explicitly constructed in \cite{ArmVic25}. Armstrong and Vicol prove anomalous diffusion and also show that the vanishing-$\kappa$ limit does not select a unique solution of the underlying, inviscid, advection equation. Therefore uncertainty in the diffusivity leads to nontrivial inviscid statistics \cite{RufSim25}. By contrast, perturbations of the initial condition for a given value of the diffusivity $\kappa$ are controlled by
\[
\|\theta^{\kappa}_1(t)-\theta^{\kappa}_2(t)\|_{L^2_x} \leq \|\theta^\kappa_1(0)-\theta^\kappa_2(0)\|_{L^2_x},
\]
which follows from incompressibility of the advecting field and from linearity of the equation. The passive scalar dynamics \cite{ArmVic25} therefore yield an example of lack of convergence toward a unique solution of the inviscid dynamics while preserving continuity with respect to initial condition. Another example in finite dimension is given by the regularization of the Peano system $\dot{x}= \sqrt{|x|}$ discussed in \cite{RufSim26}.
\end{remark*}
The question of whether or not the inviscid limit looses its continuity with respect to the initial condition might depend drastically on the initial condition and on the type of perturbation. The former dependence is left for future work while the latter can be investigated by choosing the rate of convergence of the approximation of the initial data through the choice of $\lambda$. Let us derive the expected critical value for $\lambda$ above which continuity with respect to the initial condition is expected. A phase perturbation of size $\varepsilon^\lambda$ is expected, generically, to shift the collapse time by $\Delta t^\star=O(\varepsilon^\lambda)$. As sketched in earlier sections, viscosity arrest collapse at a scale $L_\nu\sim \nu^{\frac{1}{2-\alpha}}$ and $\tau_\nu\sim \nu^{\frac{\alpha}{2-\alpha}}$ amount of time before $t^\star$, whereas saturation acts when $\sigma|\psi|^4\sim1$, giving $\tau_\sigma\sim\sigma^{1/2}$. The relative displacement in the inner collapse variables is therefore $\Delta t^\star/\tau_\varepsilon$. Assuming that an order-one displacement is retained by the post-collapse dynamics, loss of continuity is expected for
$$
\lambda<\lambda_{\rm c}^{\nu}= \frac{\alpha}{2-\alpha}
$$
in the viscous case and
$$
\lambda<\lambda_{\rm c}^{\sigma}=\frac12
$$
in the saturating case. For $\alpha=1/2$, these thresholds are respectively $1/3$ and $1/2$. This remains compatible with weak collapse, since the core mass scales as $A^2L\sim\tau^{1/\alpha-1}=\tau\to0$: the collapse amplifies phase information without carrying or dissipating a finite amount of mass. We ran simulations (not presented here) with $\lambda =1> \lambda_{\rm c}^{\varepsilon}$ showing a vanishing limiting mass gap. In the following, we will present results obtained with $\lambda=1/4< \lambda_{\rm c}^{\varepsilon}$, leading to a sustained mass gap in the limit. Although only partial, these results go in the sens of the conjectured critical value $\lambda_{\rm c}^{\varepsilon}$.
\subsection{Numerical results on post-blowup unpredictability}
\label{sec:NumLSP}

Sections~\ref{sec:LSP_reg} and \ref{sec:LCWRTIC} introduced two diagnostics of the deterministic inviscid limit. The first, $\mathcal{G}_{\varepsilon_1,\varepsilon_2}$, tests whether the regularized dynamics selects a unique weak solution of the inviscid problem after blowup. The second, $\widetilde{\mathcal{G}}_{\varepsilon}$, measures sensitivity to a vanishing random perturbation of the initial condition. Since two regularizations and two sources of uncertainty give four different cases in total, we present the viscous regularization in the main text and defer the corresponding results for the saturating regularization to Appendix~\ref{app:SaturReg}.
\begin{figure}[htb]
\centering
\includegraphics[width=0.95\linewidth]{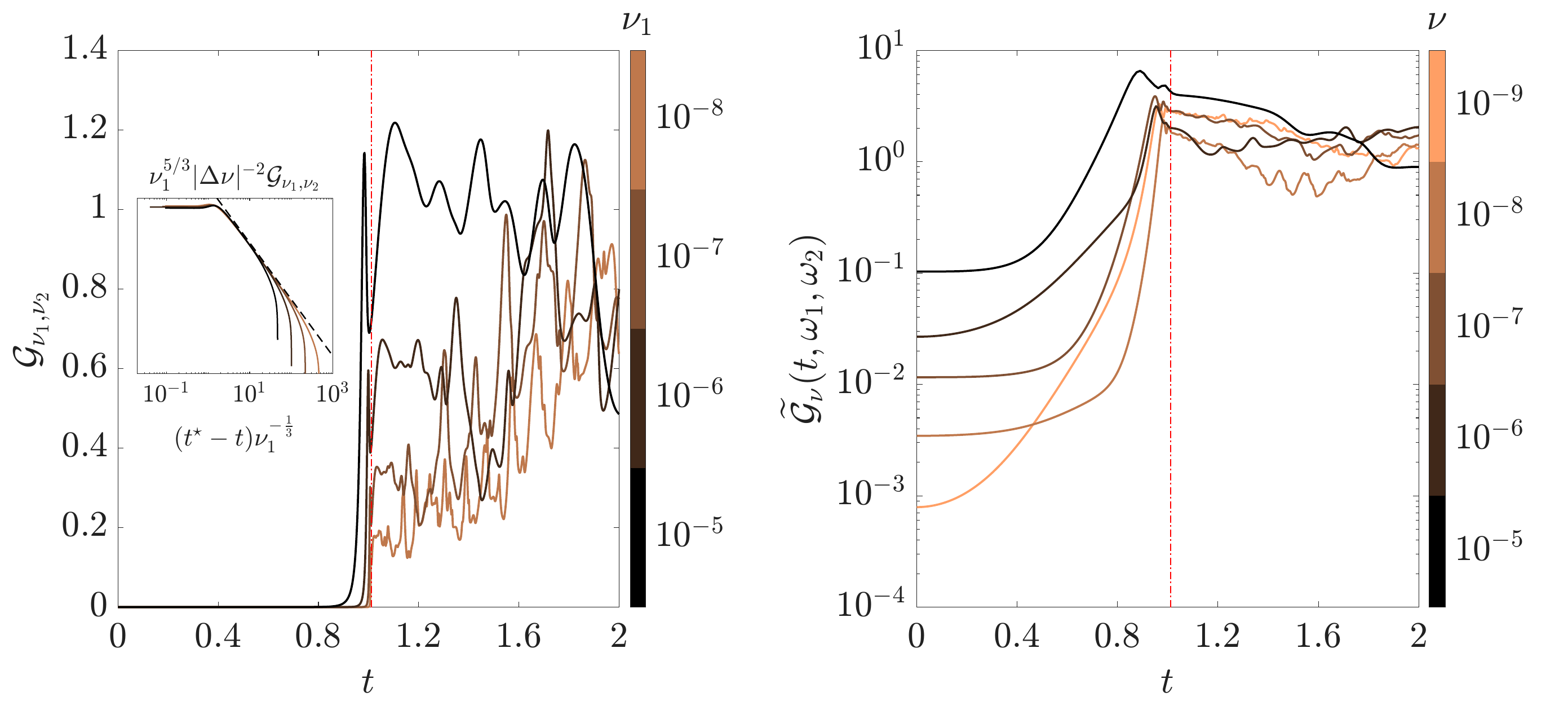}
\caption{Left: evolution of the mass gap \eqref{eq:MassErrorParamReg} for the viscous regularization. The colorbar indicates $\nu_1$. The mass gap vanishes before blowup and approaches a finite post-blowup value as the regularization parameter decreases, indicating nonselection. The inset shows the rescaled data, $\nu_1^{5/3}\Delta\nu^{-2}\mathcal{G}_{\nu_1,\nu_2}$ against $(t^\star-t)\nu_1^{-1/3}$; colors match the main plot and the black dashed line shows the scaling \eqref{eq:ScalingMassError}. Right: evolution of the mass gap \eqref{eq:MassErrorRandomInit} generated by the randomized initial condition \eqref{eq:RandomInit}. The colorbar indicates $\nu$, and the red vertical lines in both panels mark $t^\star$. Before blowup, the gap decreases as $\varepsilon^{2\lambda}$. After blowup, it remains finite in the inviscid limit. Illustrative movies show \href{https://youtu.be/fONDEtexK_4}{collapse at two different viscosities} and \href{https://youtu.be/_cadYjX16uY}{the evolution from perturbed initial conditions at fixed viscosity} (five solutions in the left panel and a Fourier mode for twenty realizations in the right panel).}
\label{fig:LSP}
\end{figure}
The left panel of Figure~\ref{fig:LSP} displays the evolution of $\mathcal{G}_{\nu_1,\nu_2}$ as a function of time for various viscosity pairs $(\nu_1,\nu_2)$. The mass gap tends to zero for $t<t^\star$, consistently with convergence to the unique smooth inviscid solution before blowup (weak-strong uniqueness). However, in order to recover a finite mass gap after blowup, the amplification of $\mathcal{G}_{\nu_1,\nu_2}$ necessarily becomes singular at $t^\star$ in the limit. In Appendix~\ref{app:PreBlowMassGap} we indeed derive such singular behavior using heuristic arguments based on the self similar collapse assumption \eqref{eq:ConjBlowupPsi} leading to
\begin{equation}
\label{eq:ScalingMassError} \lim_{\underset{\nu_1\neq \nu_2}{\nu_1,\nu_2\to 0}} |\nu_2-\nu_1|^{-2}\mathcal{G}_{\nu_1,\nu_2}(t) \underset{t\to t^\star}{\propto} (t^\star-t)^{1-\frac{3}{\alpha}}.
\end{equation}
With our choice of $\alpha=1/2$, the amplification rate is expected to be a power law of the distance to $t^\star$ with exponent $-5$. This yield the inviscid picture. Let us now present the same weak-collapse heuristic argument as earlier for anomalous dissipation. The separation between the two solutions saturates when the coarser one reaches its peak amplitude set by $\nu_1$. That is, for $t^\star-t= \tau_{\nu_1}\sim \nu_1^{\frac{\alpha}{2-\alpha}} $. At this point, following the asymptotic \eqref{eq:ScalingMassError} the mass gap reaches a plateau,
\begin{equation}
\label{eq:HeursiticMassGap} \mathcal{G}_{\nu_1,\nu_2}(t) \sim |\Delta \nu |^2 \tau_{\nu_1}^{1-\frac3 \alpha} \sim|\Delta \nu |^2 \nu_ 1^{\frac{\alpha-3}{2-\alpha}} \sim \nu_1^{\frac{1-\alpha}{2-\alpha}}
\end{equation}
Without surprise, this is the same scaling as the mass contained in a single collapsing core: in the weak-collapse regime $\alpha<1$, a single collapse event is therefore unable to create a mass gap in the limit. The inset of the left panel of Figure~\ref{fig:LSP} displays the growth of the mass gap as a function of the distance to blowup time. The amplitude of the mass gap is rescaled by the heuristic scaling \eqref{eq:HeursiticMassGap} and the distance to blowup time is rescaled by $\nu_1^{1/3}$ which is the small viscosity behavior of $\tau_{\nu_1}$ for $\alpha =1/2$. Additionally, we superimpose the power law scaling \eqref{eq:ScalingMassError} whose exponent is $-5$ for $\alpha =1/2$. This inset validate the heuristic, weak-collapse picture. The rescaled data reach a plateau independent of viscosity below the viscous time, namely for $(t^\star-t)\nu_1^{-1/3}\simeq 1$, this validates both the amplitude of this plateau \eqref{eq:HeursiticMassGap} and the scaling $\tau_\nu \sim \nu^{1/3}$. Additionally, at earlier times $t< t^\star- \tau_{\nu_1}$, all simulations follow the inviscid scaling $(t^\star-t)^{1-\frac{3}{\alpha}}$. This inset shows compelling evidence that the heuristic arguments developed here indeed constitute the leading order underlying mechanism. The interpretation of the sustained mass gap in the inviscid limit for large time is therefore the same as the one of anomalous dissipation: each collapse contribute $\sim \nu^{\frac13}$ to the total mass gap but the number of collapsing event grows as $\sim \nu^{-\frac13}$, enabling a finite mass gap in the limit.

Interestingly, the argument developed in Appendix~\ref{app:PreBlowMassGap} leading to the scaling \eqref{eq:ScalingMassError} is regularization dependent. In particular, we also discuss the case of \eqref{eq:SaturMMT} yielding an exponent $1/\alpha-5$ which is equal to $-3$ for $\alpha=1/2$. Numerical results for the saturating nonlinearity are given in Appendix~\ref{app:SaturReg} and are in accordance with a $-3$ power law exponent. It might be surprising that the pre-blowup behavior of the solution depends on the regularization mechanism. The situation is in fact subtle, when computing $|\varepsilon_1-\varepsilon_2|^{-2}\mathcal{G}_{\varepsilon_1,\varepsilon_2}(t)$ one is in fact probing the vanishing regularization behavior of $\partial_\varepsilon \Psieps$ as detailed in Appendix~\ref{app:PreBlowMassGap}. There is no a priori reason that $\partial_\varepsilon \Psieps$ should be independent of the regularization mechanism, and it is not the case in practice as exemplified by the different power law exponents.

The right panel probes sensitivity to the initial condition. It shows the evolution of $ \widetilde{\mathcal{G}}_{\nu}(t,\omega_1,\omega_2)$ for the viscous model. Before $t^\star$, the mass gap is proportional to $\nu^{2\lambda}$, as suggested by the figure. At initial time, this follows from the bound \eqref{eq:BoundRandomInit}, which is linear in $\nu^{\lambda}$, together with the quadratic nature of the mass. In the next section we show that this $\nu^{2\lambda}$ scaling persists until inviscid blowup after averaging over the realizations of the initial condition. After blowup, the mass gap no longer decreases with the regularization parameter and remains finite in the inviscid limit, consistently with the criterion $\lambda<\lambda_{\rm c}^{\nu}$ derived in Section~\ref{sec:LCWRTIC}. This observation also rules out a purely chaotic interpretation of the post-blowup dynamics. In an ordinary chaotic system, reducing the perturbation amplitude would eventually restore predictability at fixed time. Here the perturbation of the initial condition vanishes with the regularization parameter, yet the corresponding uncertainty remains macroscopic after blowup. The post-blowup dynamics is therefore not merely difficult to predict; it loses deterministic selection in the singular limit. The simulations used for the right panel were performed with independent realizations of the random phases $\theta_\omega$ for each run. The qualitative behavior shown in Fig.~\ref{fig:LSP} appears robust with respect to the realization, and should therefore persist after averaging, as discussed below. Taken together, these results lead to a clear conclusion: the inviscid limit cannot be understood in the usual deterministic sense after blowup. The limiting dynamics is neither uniquely selected by the regularization nor stable under vanishing perturbations of the initial condition. The same qualitative conclusion holds for the saturating nonlinearity, as detailed in Appendix~\ref{app:SaturReg}. The numerically accessible values of $\sigma$ do not, however, allow us to claim convergence toward a fully developed $\sigma\to 0$ regime. The results for \eqref{eq:SaturMMT} should therefore be read as qualitative evidence for the same collapse-induced loss of deterministic selection, rather than as a quantitative characterization of the vanishing-saturation limit.
\section{ Emergence of statistical order from the singular limit}
\label{sec:StatisticalSelection}

The previous section showed that the post-blowup inviscid limit does not select a single continuation. Vanishing perturbations of the regularization parameter or of the initial condition produce finite discrepancies after collapse. Following the selection framework of Section~\ref{sec:SpTrip}, we now investigate the selection of statistics. We consider two vanishing randomizations. The first randomizes the regularization parameter in an interval shrinking to zero. The second randomizes the initial condition by a perturbation whose amplitude also tends to zero. As a consequence of the lack of deterministic selection presented in the previous section, the vanishing randomness will be amplified to a macroscopic level upon wave collapse. This allows to interpret the inviscid limit as a genuinely random process.

In the following sections, we study the inviscid limit as a statistical limit rather than a deterministic one. We begin in Section~\ref{sec:NoiseMechanisms} with bounds on the fluctuations induced by these randomizations, proved in Appendix~\ref{app:AnomFluc}. They show that, for both randomization mechanisms, the fluctuations vanish in the inviscid limit as long as the solution remains regular, as the uniqueness of the strong solution before blowup leads one to expect. We then present, in Section~\ref{sec:StatInvLimit}, the numerical results on the statistical description of the inviscid limit.
\subsection{Randomization mechanisms and pre-blowup determinism}
\label{sec:NoiseMechanisms}

We distinguish the regularization mechanism, indexed by \(\varepsilon\in\{\nu,\sigma\}\), from the source of randomness, indexed by \(\chi\in\{\mathbf r,\mathbf i\}\). The index \(\mathbf r\) denotes randomization of the regularization parameter, while \(\mathbf i\) denotes randomization of the initial condition. In what follows we randomize the initial condition and the regularization parameter separately. Randomizing both at once is an interesting question, which we leave for future work. The two randomization are

\begin{equation}
\label{eq:RandInitCond} \psi_0^{\varepsilon,\omega}(x) = \sum_{k\in\mathbb{Z}} e^{i\varepsilon^{\lambda}\theta^{\mathbf{i}}_\omega(k)} \widehat{\psi}_0(k)e^{i kx}, \qquad \varepsilon(\omega) = \varepsilon + \varrho_{\varepsilon}^{\mathbf r}(\omega).
\end{equation}

Here \(\theta^\mathbf{i}_\omega(k)\) are independent random phases and $\lambda>0$ is the exponent introduced in Section~\ref{sec:LCWRTIC} controlling how fast the random initial condition converges toward the deterministic one \eqref{eq:InitCond}. The distribution of \(\varrho^{\mathbf r}_{\varepsilon}\) is centered and chosen so that \(\varepsilon(\omega)\ge0\) for every realization and \(\varrho^{\mathbf r}_{\varepsilon}(\omega)\to0\) uniformly in $\omega$ as \(\varepsilon\to 0\). In both randomizations of \eqref{eq:RandInitCond}, the parameter $\varepsilon$ is deterministic: it is the regularization parameter itself when the initial condition is randomized, and the mean regularization parameter when the regularization is randomized. Following \ref{it:spst-statistical}, the solution $\psi_{\varepsilon,\chi}^\omega$ of \epsfNLSref{} with randomization $\chi$ at fixed time is then viewed as a random variable on the space of finite mass fields, with law given by the corresponding pushforward measure. In other words, the randomizations \eqref{eq:RandInitCond} give a well-defined probabilistic meaning to the regularized problem. We are then interested in characterizing the probability distribution of the solution, with respect to either source of randomness, in that limit. This distribution lives on the infinite-dimensional space of square-integrable functions, so a complete statistical description of the limit is out of reach, and we focus instead on lower-dimensional signatures of a possible statistical selection. The first is the variance of the distribution in physical space: a variance that does not vanish in the inviscid limit is strong evidence that randomness persists.

Denoting by $\mathbb{E}_\chi$ the average with respect to the randomization $\chi\in\lbrace \mathbf{i},\mathbf{r} \rbrace$, the variance of the law of the regularized solution reads
\begin{equation}
\label{eq:DefFlucMass} \mathsf{M}_{\varepsilon,\chi }(t,x):=\mathbb E_\chi{\left|\psi^\omega_{\varepsilon,\chi}(t,x)-\Echi{\psi^\omega_{\varepsilon,\chi}(t,x)}\right|^2}, \quad \mathsf{M}_{\varepsilon,\chi }(t):=\int_{\T} \mathsf{M}_{\varepsilon,\chi }(t,x) \,dx,
\end{equation}
where $\psi^\omega_{\varepsilon,\chi}$ is the solution of the regularized dynamics \epsfNLSref{} with random regularization parameter $\varepsilon(\omega)$ and deterministic initial condition $\psi_0$ given by \eqref{eq:InitCond} if $\chi=\mathbf{r}$ or the solution of \epsfNLSref{} with deterministic regularization parameter $\varepsilon$ and random initial condition $\psi_0^{\varepsilon,\omega}$ given by \eqref{eq:RandInitCond} if $\chi=\mathbf{i}$. We say that the dynamics exhibits anomalous fluctuations if these quantities remain positive in the inviscid limit. Note that the mass of the fluctuation field \eqref{eq:DefFlucMass} is directly related to the mass gaps introduced above. From the definition of the fluctuation field and for two independent realizations $(\omega_1,\omega_2)$ of the randomization mechanism,
\begin{equation}
\label{eq:TwoCopyFluc} \mathbb{E}_{\mathbf i}^{(\omega_1,\omega_2)} \left[\widetilde{\mathcal{G}}_{\varepsilon}(t,\omega_1,\omega_2)\right] = 2\mathsf{M}_{\varepsilon,\mathbf i}(t), \qquad \mathbb{E}^{(\omega_1,\omega_2)}_{\mathbf r} \left[ \mathcal{G}_{\varepsilon(\omega_1),\varepsilon(\omega_2)}(t)\right] = 2\mathsf{M}_{\varepsilon,\mathbf r}(t).
\end{equation}
Thus the persistent mass gaps observed after blowup in Figure~\ref{fig:LSP} are precisely the deterministic signature that a vanishingly small amount of uncertainty either on the initial condition or on the regularization parameter can produce a macroscopic effect on the solution, after blowup. Before blowup however, the inviscid solution is smooth and unique, mass flucutations must therefore vanish in the inviscid limit. This pre-blowup control on fluctuations is precisely stated in Proposition~\ref{prop:PreBlowFluctuations} in Appendix~\ref{app:AnomFluc} and in particular we have the following bounds for $t\in[0,T], \; T<t^\star$,
\begin{align}
\mathsf M_{\nu,\mathbf i}(t) &\leq C\nu^{2\lambda}\mathcal{M}_0 \exp\left[C\int_0^t\left(1+\|\psi(s)\|_{L^\infty}^2\right)ds\right], \label{eq:PreBlowFlucInit}\\
\mathsf M_{\nu,\mathbf r}(t) &\leq C\nu^2\left[\int_0^t\left(1+\|\Delta\psi(s)\|_{L^2}^2\right)ds\right] \exp\left[C\int_0^t\left(1+\|\psi(s)\|_{L^\infty}^2\right)ds\right]. \label{eq:PreBlowFlucVisc}
\end{align}
The constant $C$ is independent of $\nu$ and $T$, $\mathcal{M}_0$ is the mass of the initial condition. The bounds \eqref{eq:PreBlowFlucInit} and \eqref{eq:PreBlowFlucVisc} are instructive in two ways. First, they show that as long as the solution $\psi$ of the inviscid problem \eqref{eq:MMT} is smooth, the joint limit of vanishing regularization and vanishing noise, $\nu\to0$, forces the variance of the law of the solution to vanish: the solution is deterministic. Second, they show how this picture breaks down, since both involve the squared $L^\infty$ norm of the inviscid solution, which is precisely the quantity that diverges at blowup. The numerical results of Figure~\ref{fig:BlowupAnomDiss} suggest that $\|\psi\|_{L^\infty}^2\propto(t-t^\star)^{-1}$, the borderline case for which the bound ceases to be finite, and which therefore leaves open the possibility that randomness be amplified to large scales. In the weak-collapse regime considered here at $\alpha=1/2$, however, the heuristic argument \eqref{eq:HeursiticMassGap}, validated numerically in Figure~\ref{fig:LSP}, indicates that a single collapse cannot produce finite mass fluctuations. The bounds \eqref{eq:PreBlowFlucInit} and \eqref{eq:PreBlowFlucVisc} are therefore not sharp: the proof given in Appendix~\ref{app:AnomFluc} retains only the peak amplitude of the solution and not its width. Combining the two would be required to establish rigorously that a single collapse of the mass-supercritical ($\alpha<1$) fractional NLS \eqref{eq:MMT} cannot generate finite mass fluctuations.

Informative as it is, such a variance does not by itself establish convergence of the law. As a partial answer to the question of statistical selection, we therefore also compute the probability laws of individual Fourier modes and examine whether they converge toward a non-Dirac distribution. We now turn to the numerical results on the statistical description of the inviscid limit.
\subsection{Statistical description of the limit }
\label{sec:StatInvLimit}

\begin{figure}[htb]
\centering
\includegraphics[width=0.94\linewidth]{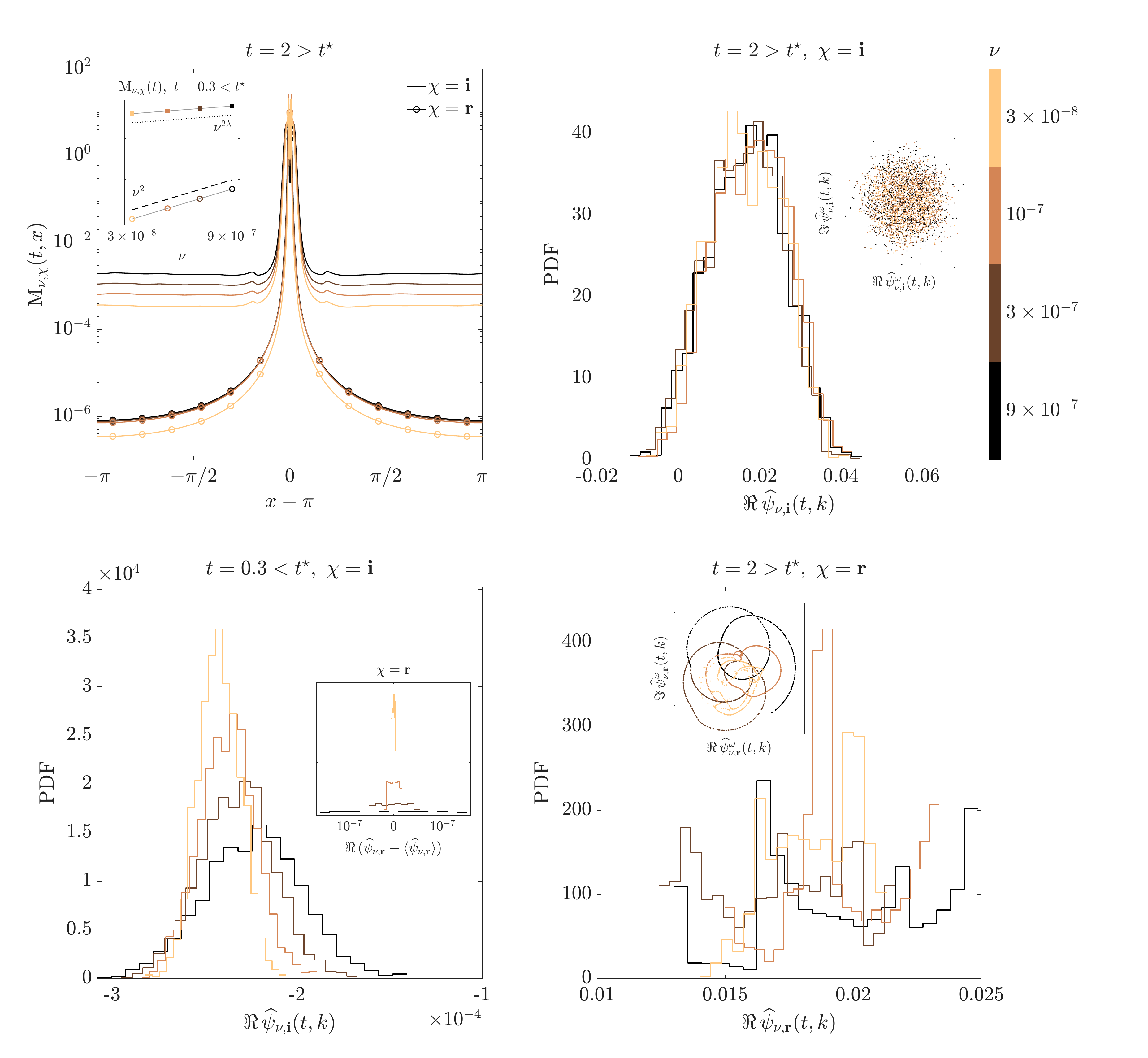}
\caption{Statistical descirpition of the inviscid limit for the viscous regularization $\varepsilon = \nu\in \lbrace 0.3,1,3,9 \rbrace\times 10^{-7}$ and randomization of initial condition $\chi=\mathbf{i}$ and regularization parameter $\chi=\mathbf{r}$. In the main figure of the upper left panel, we display the local fluctuation of mass $\mathsf{M}_{\nu, \chi}(t,x)$ at $t=2>t^\star$ for both randomization mechanisms (solid lines for random initial condition and circle markers for random viscosity) and for decreasing values of the viscosity stated in the colorbar. While the curves in the main panel shows finite local mass fluctuations in the limit, the inset displays the behavior of the integrated fluctuation mass $\mathsf{M}_{\nu, \chi}(t)$ for the pre-blowup time $t= 0.3 <t^\star$. Both dotted lines represent the viscosity dependence of the bounds \eqref{eq:PreBlowFlucInit} \& \eqref{eq:PreBlowFlucVisc}. In the bottom left panel, we display the viscosity dependence of the probability distribution of $\Re \, \widehat{\psi}_{\nu,\chi}(t,k)$ in the pre-blowup state $t= 0.3 < t^\star$, for $k= 64$ and $\chi= \mathbf{i}$ in the main figure and $\chi= \mathbf r$ in the inset. For both regularization, the pre-blowup distribution converges toward a Dirac mass. The upper and lower right panels display the post-blowup distributions for $t=2 > t^\star$ and $k=64$ and for the four viscosities. The upper right panel is dedicated to the randomized initial condition $\chi= \mathbf i$ and the lower-right panel to the random viscosity $\chi= \mathbf r$. In both cases, the insets displays $800$ independent samples for each value of the viscosity in the complex plane.}
\label{fig:AnomFluc}
\end{figure}

Finite variance cannot survive before the inviscid blowup, by \eqref{eq:PreBlowFlucInit} and \eqref{eq:PreBlowFlucVisc}, and cannot be sustained by a single collapse. We now ask what becomes of a vanishingly small randomness after blowup, in the collapse-proliferation scenario. We report the behavior of the fluctuation mass $\mathsf{M}_{\nu,\chi}$ and the probability distribution of $\widehat{\psi}^\omega_{\nu,\chi}(t,k)$ for both randomizations \eqref{eq:RandInitCond}, for the viscous regularization only; the corresponding results for the saturating nonlinearity are given in Appendix~\ref{app:SaturReg}. We recall that the initial condition and regularization parameter are randomized according to \eqref{eq:RandInitCond}. In the numerical simulations, we consider the following sequences for regularization parameters
\[
(\nu_{\mathbf{r}}^1,\nu_{\mathbf{r}}^2,\nu_{\mathbf{r}}^3,\nu_{\mathbf{r}}^4)=\left(9,3,1,0.3\right)\times 10^{-7}, \qquad (\sigma_{\mathbf{r}}^1,\sigma_{\mathbf{r}}^2,\sigma_{\mathbf{r}}^3,\sigma_{\mathbf{r}}^4)=\left(5,\frac{5}{\sqrt{2}},\frac{5}{2},\frac{5}{2\sqrt{2}}\right)\times 10^{-4}.
\]
For randomized regularization parameters, we take
\begin{equation}
\label{eq:ParamRandom} \nu(\omega)=\nu_{\mathbf{r}}\bigl(1+\dfrac{1}{2}\eta(\omega)\bigr), \qquad \sigma(\omega)=\sigma_{\mathbf{r}}\bigl(1+(1-\dfrac{1}{\sqrt{2}})\vartheta(\omega)\bigr),
\end{equation}
where $\eta$ and $\vartheta$ are uniformly distributed on $[-1,1]$. All averages are computed using $2\times 10^3$ independent samples and probability distributions are estimated using histrograms. We display in Figure~\ref{fig:AnomFluc} the statistical study of the limit.

In the upper left panel of Figure~\ref{fig:AnomFluc} we study the second statistical order of the limit in physical space. The main panel displays the local mass fluctuations $\mathsf{M}_{\nu,\chi}(t,x)$ at the final simulation time $t=2>t^\star$, for vanishingly small viscosities and for both randomization mechanisms. In both cases the mass fluctuations survive in the inviscid limit. Moreover, at the available viscosities $\nu\in\lbrace 0.3,1,3,9\rbrace\times10^{-7}$, the local fluctuations appear to converge toward a well-defined profile, which depends on the randomization mechanism and is localized in a bulk region around the first collapse at $x=\pi$. Notably, although each realization of the randomized initial condition \eqref{eq:RandInitCond} breaks the spatial symmetry about $x=\pi$, that symmetry appears to be restored at the statistical level. The inset of the same panel displays the space-integrated mass fluctuations $\mathsf{M}_{\nu,\chi}(t)$ at the pre-blowup time $t=0.3<t^\star$, for each randomization, $\chi=\mathbf{i}$ (circles) and $\chi=\mathbf{r}$ (squares), together with the corresponding powers of the viscosity obtained in the bounds \eqref{eq:PreBlowFlucInit} and \eqref{eq:PreBlowFlucVisc}, drawn as black dashed lines. The numerical data are in excellent agreement with these rates, confirming that the fluctuations vanish before the first inviscid blowup.

Note however that although it is a good indicator for determinism, a vanishing variance does not strictly speaking mean that the preblowup distribution converges toward a Dirac mass\footnote{Take for example a sequence of centered Gaussian random variables $X_\epsilon$ of variance $\sigma^2_\epsilon= \ln 1/\epsilon$ and define $M_\epsilon = \epsilon^{q}e^{X_\epsilon}$ for some $q>0$. Then for $p>0$, $\mathbb{E}[M_\epsilon^p]=\epsilon^{p(q-p/2)}$. In particular, whenever $q>1$, the field has vanishing variance but infinite higher order moments in the limit $\epsilon \to 0$. }. Thereby, we also investigate the probability distribution of the real part of the Fourier mode $k=64$ of the solution, before and after blowup, for each randomization mechanism and the four values of the viscosity. The pre-blowup distribution of $\Re \widehat{\psi}_{\nu, \chi}(t,k)$ is shown in the bottom left panel of Figure~\ref{fig:AnomFluc}. The main panel displays the case $\chi=\mathbf i$ and the inset the case $\chi=\mathbf r$. We checked (not shown) that both randomization converge toward the same Dirac mass. For readibility purposes however, we cannot display the distritution on the same panel since their variance differs by three orders of magnitude and we show the centered variable in the $\chi= \mathbf r$ for the same readability purpose. For both sources of randomness, the distribution becomes more sharply peaked as the viscosity vanishes. Pre-blowup predictability is thus recovered, the limiting distributions being Dirac masses independent of $\chi$. After blowup, by contrast, randomness survives, as the upper left panel already indicates. At $t=2>t^\star$, the upper right panel suggests that the distribution of $\Re \widehat{\psi}_{\nu, \mathbf{i}}(t,k)$ converges toward a non-Dirac law; other Fourier modes converge as well, toward different laws (not shown). The distribution of $\Re \widehat{\psi}_{\nu, \mathbf{r}}(t,k)$ represented in the bottom right panel, by contrast, shows no clear sign of convergence as the regularization is removed. These two limiting behaviors illustrate the measure-theoretic definition of spontaneous stochasticity introduced in \cite{RufSim26} and recalled in Section~\ref{sec:SpTrip}. The inviscid statistics depend in an intricate way on the inviscid problem \ref{it:spst-inviscid}, on the regularization procedure \ref{it:spst-regularized} (see Appendix~\ref{app:SaturReg} for the discussion of \eqref{eq:SaturMMT}), and on the source of randomness \ref{it:spst-statistical}. This last dependence is plainly illustrated by the two sources used here. When the uncertainty lies on the initial condition alone, the probability distribution seems to converges in the limit, so that a well-defined statistical meaning can be given to the inviscid limit where the deterministic one fails; this is strong spontaneous stochasticity in the sense of \cite{RufSim26}. When the regularization parameter alone is sampled, the distribution shows no sign of convergence. A lack of convergence of the distribution in the limit means that the inviscid limit can neither be understood deterministically, nor statistically.

Beyond the distributions of $\Re \widehat{\psi}_{\nu, \chi}(t,k)$, we also show in the insets of the upper- and lower right panels $800$ independent samples of $\widehat{\psi}^\omega_{\nu, \chi}(t,k)$ in the complex plane for each value of the (average) regularization parameter $\nu$. The qualitative difference between the two sample distributions is striking. It stems from the dimensionality of the random perturbation. For a random viscosity ($\chi=\mathbf{r}$), varying a single real parameter traces a curve $\nu\mapsto\widehat{\psi}_{\nu}(t,k)$ in the complex plane, as the inset shows, and the distribution is therefore concentrated along that curve. In the vanishing-viscosity limit this curve may become space-filling, so that a limiting distribution, should it exist, could still have a density. Randomizing the initial condition, by contrast, amounts to sampling the map $\psi(0,\cdot)\mapsto\widehat{\psi}_{\nu}(t,k)$ defined on an infinite-dimensional function space, which allows the Fourier coefficient to explore a two-dimensional region of the complex plane and produces a smoother distribution.

All in all, the pre-blowup inviscid limit retains a deterministic meaning, as expected from weak-strong uniqueness before blowup. After blowup the picture is less clear-cut. When the initial condition alone is sampled, the simulations show convergence of the probability distribution, and the consequence is strong: the inviscid limit cannot be read as a deterministic solution of \eqref{eq:MMT}, but it can be read as a genuinely, spontaneously random one. This is the strong spontaneous stochasticity scenario introduced in Section~\ref{sec:SpTrip} and developed in \cite{RufSim26}. When the viscosity is sampled instead, not even the probability distribution shows a sign of convergence, although convergence at smaller viscosities cannot be excluded; this would be the signature of weak spontaneous stochasticity. The absence of convergence may also reflect the particular sampling \eqref{eq:ParamRandom} we have chosen. In the spirit of the construction of \cite{RufSim26}, which gives a statistical meaning to a non-converging sequence of regularized solutions $\psi_\varepsilon$, one could then seek a statistical meaning for a non-converging sequence of laws and describe the inviscid limit by a probability measure on the space of probability measures. What the present simulations do establish is that a vanishing source of randomness can leave a finite and reproducible statistical effect.

\section{Conclusion and perspectives}

We have studied a focusing fractional nonlinear Schr\"odinger equation in a regime where the inviscid dynamics undergoes finite-time wave collapse. The guiding question was whether such a singularity can generate, in a dispersive medium, the same kind of singular-limit phenomena that are usually associated with fluid turbulence. Our results suggest a subtle route toward anomalous dissipation and spontaneous stochasticity. The studied equation is in the so-called mass supercritical regime, also known as weak collapse. In such a regime, a collapsing wave carries a vanishing amount of mass. One can therefore legitimately doubt that such a model could exhibit anomalous dissipation and spontaneous stochasticity. However, after the first inviscid collapse, the regularized dynamics undergoes several posterior singular event, whose number diverges in the inviscid limit. The vanishingly small mass contained in collapse cores is exactly compensated by their proliferation in the limit, yielding both anomalous dissipation and lack of selection of a single inviscid solution in the limit.

We first compared two collapse-arresting mechanisms, viscous diffusion and nonlinear saturation. For each fixed value of the regularization parameter, both mechanisms prevent blowup. As the regularization is removed, both recover the same smooth inviscid solution before the first collapse time. After collapse, however, they select different weak continuations. The vanishing viscosity limit allows for mass dissipation, giving a dispersive analogue of anomalous dissipation, whereas the saturating limit remains conservative. Thus the post-blowup mass balance is not determined by the inviscid equation alone. It depends on the mechanism used to arrest collapse.

The coarse-grained mass balance clarifies how this can happen. For smooth solutions, the nonlinear commutator responsible for the mass defect vanishes. This remains true under sufficient integrability, in particular under \(L^4\) control. Wave collapse destroys precisely this control. The post-blowup regime is therefore flexible enough to accommodate different weak mass balances, in direct analogy with Onsager-type conservation defects in fluid turbulence.

We then tested whether deterministic selection is restored once the regularization mechanism is fixed. Our simulations indicate that it is not. Two solutions obtained with different but arbitrarily close regularization parameters separate in finite time. The same behavior is observed for perturbations of the initial condition whose size vanishes with the regularization parameter. The post-blowup limit is therefore neither uniquely selected within a fixed regularization class nor continuous with respect to the initial state. This differs from ordinary chaotic sensitivity. The uncertainty is not merely amplified over long times. It survives at finite time in the singular inviscid limit.

Finally, we replaced this failed deterministic selection by a statistical one. Introducing a source of randomness vanishing with the regularzation parameter allows to give a well defined statistical meaning to the regularized problem. In the inviscid limit, even though randomness vanishes its effect can survive and produce macroscopic effects and lead a well-defined probabilistic meaning of the limit. To study this, we sample either the regularization parameter or the initial condition. We find that randomizing the initial condition leads to a well-defined statistical meaning for the limit. The tryptich inviscid model \eqref{eq:MMT}, viscous or saturating regularization \epsfNLSref{} and vanishing uncertainty on the initial condition \eqref{eq:RandInitCond} is Spontaneously Stochastic. However, when the initial condition is taken deterministic but the regularization parameter fluctuates, there are no clear sign of convergence of the probability distribution. This suggests that in this case, even a statistical description of the limit fails.

These results identify collapsing dispersive equations as a testbed for singular inviscid limits in analogy with fluid turbulence. In both cases, the singular limit may carry an anomalous conservation defect, and the limiting dynamics may be statistical rather than deterministic. The mechanism is nevertheless different. In the present model, uncertainty is produced by finite-time wave collapse rather than by advective stretching across an inertial range. In the terminology of Section~\ref{sec:SpTrip}, the present results show that the limiting law is not a Dirac mass. Depending on how the inviscid limit is sampled (i.e if one considers random initial condition or random regularization parameter ), the results suggest that one can obtain a well-defined statistical limit or not.

A first one is to test the robustness of the observed statistical state with respect to parameters kept fixed here, including the initial condition, the dispersion exponent, the initial mass. A second direction is to move closer to experimentally relevant optical systems, in particular by considering two-dimensional NLS dynamics and saturation or damping laws more realistic of nonlinear optics. In this respect, nonlinear optical media appear as promising candidates for observing finite-time loss of deterministic selection in a controlled dispersive setting. We note however that the two dimension cubic NLS is mass critical. In that regard, wave collapse carries a finite amount of mass. Collapse proliferation is therefore exluced in this case. A possible route toward Spontaneous Stochasticity in such systems is however the stochastic post-collapse beam interaction studied by Fibich and collaborators. See \cite{Fib15} and references therein. Finally, it would be interesting to investigate dispersive models where ill-posedness does not originate from focusing collapse. Possible examples include energy-supercritical defocusing NLS, where finite-time blowup mechanisms closer to compressible Euler dynamics have recently been constructed \cite{MerRap22}, and rough-data mechanisms such as norm inflation \cite{RufTzv26}.
\section*{Acknowledgements}

The author thanks ENS de Lyon for funding and m{\'e}socentre CBPSMN for numerical resources. The author also thanks E. Simonnet for helpful comments and for the thorough work in \cite{RufSim25,RufSim26}, on which the present article strongly relies. The author finally thanks L. Chevillard, N. Tzvetkov, and N. Valade for their decisive comments concerning the development of this manuscript. The author also thanks the anonymous referees for their suggestions and remarks that substantially improved this manuscript.

\appendix
\section{Regularizations prevent blowup}
\label{app:GWP}

We state global well-posedness for the periodic regularized equations, together with the bounds used in the main text. The Sobolev regularity assumed below is sufficient for a direct local theory; no claim of optimal regularity is made.
\begin{proposition}[Global well-posedness of the regularized equations]
\label{prop:GWP} Let $\T=\R/(2\pi\mathbb Z)$, let $\alpha>0$, and let $s\geq\max\{1,\alpha/2\}$. Suppose that $\psi_0\in H^s(\T;\mathbb C)$ and set \( \mathcal M_0=\|\psi_0\|_{L^2(\T)}^2\).
\begin{enumerate}[label=\textnormal{(\roman*)}]
\item For every $\sigma>0$, the saturating equation \eqref{eq:SaturMMT} has a unique global mild solution $\Psisig\in C(\R;H^s(\T))$. It conserves mass and the Hamiltonian \eqref{eq:HamilSatur}, and, for every $t\in\R$,
\begin{equation}
\label{eq:GWPSaturBounds}
\begin{aligned}
\M{\Psisig(t)}=\mathcal M_0,\quad \mathcal H_\sigma[\Psisig(t)]=\mathcal H_\sigma[\psi_0],\quad\|\Lambda^{\alpha/2}\Psisig(t)\|_{L^2(\T)}^2 \leq \|\Lambda^{\frac \alpha 2}\psi_0\|_{L^2}^2+\frac{\mathcal M_0}{2\sqrt\sigma}.
\end{aligned}
\end{equation}
\item If $\alpha\geq1/2$, then for every $\nu>0$ the viscous equation \eqref{eq:ViscousMMT} has a unique global mild solution
\[
\Psinu\in C([0,\infty);H^s(\T)) \cap L^2_{\mathrm{loc}}([0,\infty);H^{s+1}(\T)).
\]
For every $t\geq0$, it satisfies
\begin{equation}
\label{eq:GWPViscMass} \M{\Psinu(t)}+2\nu\int_0^t\|\partial_x\Psinu(\tau)\|_{L^2(\T)}^2\,d\tau =\mathcal M_0
\end{equation}
and
\begin{equation}
\label{eq:GWPViscBounds}
\begin{aligned}
\|\Lambda^{\alpha/2}\Psinu(t)\|_{L^2(\T)}^2 +\nu\int_0^t\|\Lambda^{\alpha/2+1}\Psinu(\tau)\|_{L^2(\T)}^2\,d\tau\leq \|\Lambda^{\frac \alpha 2} \psi_0 \|^2_{L^2(\T)}\exp\!\left(\frac{C_\alpha\mathcal M_0^2t}{\nu}\right),
\end{aligned}
\end{equation}
where $C_\alpha>0$ depends only on $\alpha$ and the fixed torus, not on $\psi_0$, $\nu$, or $t$. In particular, the estimate includes $\alpha=1/2$.
\end{enumerate}
For either equation, with its regularization parameter fixed, the solution depends continuously on $\psi_0$ in $H^s$, uniformly on every compact time interval in its stated time domain. The solutions are also $C^1$ in time with values in $H^{s-\alpha}$ for saturation and $H^{s-\max\{\alpha,2\}}$ for viscosity, and satisfy their equations in these spaces. Smooth initial data yield solutions that remain smooth for all such times.
\end{proposition}
For completeness, the local theory follows by a fixed point argument in $C([0,T];H^s)$ using the Fourier propagators $e^{-it\Lambda^\alpha}$ and $e^{t(\nu\Delta-i\Lambda^\alpha)}$. They are respectively unitary and contractive on $H^s$. Since $s\geq1$, $H^s(\T)$ is an algebra and both nonlinearities are locally Lipschitz on $H^s$. The resulting maximal solutions are unique, depend continuously on the initial data, and can be continued as long as their $H^s$ norms remain bounded. We prove the estimates first for smooth data and then pass to $H^s$ data by approximation and these stability estimates.

\subsection{Saturation prevents blowup}
\label{sec:NoBlowSatur}

\textit{Proof of Proposition~\ref{prop:GWP}\textnormal{(i)}} With the saturating nonlinearity, the dynamics conserves
\begin{equation}
\label{eq:HamilSatur} \mathcal{M}[\Psisig]= \int_\T |\Psisig(t,x)|^2\,dx, ~~~ \mathcal{H}_\sigma [\Psisig]= \int_\T \frac12\left|\Lambda^{\alpha/2} \Psisig \right|^2- \dfrac{1}{4\sigma} \ln\left(1+\sigma |\Psisig|^4 \right) ~dx
\end{equation}
From the conservation of $\mathcal{H}_\sigma$, we obtain
$$
\|\Lambda^{ \frac \alpha2} \Psisig \|^2_{L^2} = 2\mathcal{H}_\sigma[\psi_0] + \dfrac{1}{2\sigma} \int_\T \ln \left(1+\sigma |\Psisig|^4 \right) dx.
$$
Using conservation of mass and $\ln(1+x)\leq \sqrt{x}$, we obtain
$$
\|\Lambda^{ \frac \alpha2} \Psisig \|^2_{L^2} \leq 2\mathcal{H}_\sigma[\psi_0] + \dfrac{1}{2\sqrt{\sigma}} \mathcal{M}[\psi_0].
$$
Moreover, $\mathcal{H}_\sigma[\psi_0]\leq\frac12\|\Lambda^{\alpha/2}\psi_0\|_{L^2}^2$, which proves \eqref{eq:GWPSaturBounds} on the interval of existence. The conservation identities follow by differentiating the mass and Hamiltonian along smooth solutions; approximation gives the same identities for the regularity in the proposition. To prove that the solution exists for all times, we also bound its first spatial derivative. Differentiating the nonlinear term gives
\[
\left|\partial_x\left(\frac{|\Psisig|^2\Psisig}{1+\sigma|\Psisig|^4}\right)\right| \leq \frac{3|\Psisig|^2}{1+\sigma|\Psisig|^4}|\partial_x\Psisig| \leq \frac{3}{2\sqrt\sigma}|\partial_x\Psisig|.
\]
Then, differentiating the equation in $x$, multiplying by $\overline{\partial_x\Psisig}$ and taking the imaginary part yields
\[
\left|\frac{d}{dt}\|\partial_x\Psisig\|_{L^2}^2\right| \leq \frac{3}{\sqrt\sigma}\|\partial_x\Psisig\|_{L^2}^2.
\]
Using Gronwall's lemma and conservation of mass, we obtain
\begin{equation}
\label{eq:SaturHoneContinuation} \|\Psisig(t)\|_{H^1}\leq\|\psi_0\|_{H^1}\exp\left(\frac{3|t|}{2\sqrt\sigma}\right).
\end{equation}
Thus the $H^1$ norm stays finite on every finite time interval, and the local solution extends to all positive and negative times. For $s>1$, this bound also controls $\|\Psisig\|_{L^\infty}$. The usual product and chain estimates applied to the smooth saturating nonlinearity then give
\[
\left|\frac{d}{dt}\|\Psisig\|_{H^s}^2\right|\leq C_{s,\sigma,T}\|\Psisig\|_{H^s}^2,\qquad |t|\leq T,
\]
where the constant depends on the initial $H^1$ norm but not on higher norms of the solution. Another application of Gronwall's lemma therefore prevents blowup in $H^s$. Uniqueness and continuous dependence follow from the local theory stated above, completing the proof of global well-posedness.

\subsection{\texorpdfstring{Diffusion prevents blowup for $\alpha\geq\frac12$}{Diffusion prevents blowup for alpha >= 1/2}}
\label{sec:NoBLowDiff}

\textit{Proof of Proposition~\ref{prop:GWP}\textnormal{(ii)}} To establish the regularizing effect of diffusion on the torus $\T$, we first recall the mass budget
$$
\dfrac{d}{dt}\|\Psinu\|_{L^2(\T)}^2= -2\nu\|\Lambda\Psinu\|_{L^2(\T)}^2,\qquad\int_0^t\|\Lambda\Psinu(s)\|_{L^2(\T)}^2\,ds= \dfrac{1}{2\nu}\left[\|\psi_0\|_{L^2(\T)}^2-\|\Psinu(t)\|_{L^2(\T)}^2\right]\leq \dfrac{1}{2\nu}\|\psi_0\|_{L^2(\T)}^2.
$$
Then, by multiplying the diffusive dynamics \eqref{eq:ViscousMMT} by $\overline{\Lambda^\alpha\Psinu}$ and taking the imaginary part one obtains
\begin{equation}
\label{eq:SobolBudget} \dfrac{d}{dt}\|\Lambda^{\frac\alpha2}\Psinu\|_{L^2(\T)}^2+2\nu\|\Lambda^{\frac\alpha2+1}\Psinu\|_{L^2(\T)}^2=-2\Im\int_\T\overline{\Lambda^{\frac\alpha2}\Psinu}\,\Lambda^{\frac\alpha2}\left(|\Psinu|^2\Psinu\right)\,dx.
\end{equation}
H{\"o}lder's inequality on the right-hand side of \eqref{eq:SobolBudget} yields
$$
\left|\int_\T\overline{\Lambda^{\frac\alpha2}\Psinu}\,\Lambda^{\frac\alpha2}\left(|\Psinu|^2\Psinu\right)\,dx\right|\lesssim\|\Lambda^{\frac\alpha2}\Psinu\|_{L^4(\T)} \left\|\Lambda^{\frac\alpha2}\left(|\Psinu|^2\Psinu\right)\right\|_{L^{\frac43}(\T)}.
$$
Then, using the periodic Kato-Ponce inequality twice \cite[Proposition~1]{BenOhZha25}, we obtain
$$
\left\|\Lambda^{\frac\alpha2}\left(|\Psinu|^2\Psinu\right)\right\|_{L^{\frac43}(\T)}\lesssim\|\Psinu\|_{L^4(\T)}^2\|\Lambda^{\frac\alpha2}\Psinu\|_{L^4(\T)}.
$$
At this point, we are left with
$$
\dfrac{d}{dt}\|\Lambda^{\frac\alpha2}\Psinu\|_{L^2(\T)}^2+2\nu\|\Lambda^{\frac\alpha2+1}\Psinu\|_{L^2(\T)}^2\leq C\|\Psinu\|_{L^4(\T)}^2\|\Lambda^{\frac\alpha2}\Psinu\|_{L^4(\T)}^2.
$$
To use the effect of diffusion, we want to express the right-hand side in terms of $\|\Lambda^{\frac\alpha2+1}\Psinu\|_{L^2(\T)}^2$. To do so, we use the periodic Gagliardo-Nirenberg inequalities for both terms of the right-hand side, obtained from the Sobolev inequalities of \cite{BenOh13} and Fourier interpolation. This yields for $\alpha\geq 1/2$,
\begin{equation*}
\begin{cases}
\displaystyle \|\Psinu\|_{L^4(\T)}\lesssim\|\Lambda^{\frac\alpha2}\Psinu\|_{L^2(\T)}^{\frac{1}{2\alpha}}\|\Psinu\|_{L^2(\T)}^{1-\frac{1}{2\alpha}}+\|\Psinu\|_{L^2(\T)},\\[1em]
\displaystyle \|\Lambda^{\frac\alpha2}\Psinu\|_{L^4(\T)}\lesssim\|\Lambda^{\frac\alpha2+1}\Psinu\|_{L^2(\T)}^{\frac14}\|\Lambda^{\frac\alpha2}\Psinu\|_{L^2(\T)}^{\frac34}.
\end{cases}
\end{equation*}
The additional term in the first inequality accounts for the zero Fourier mode. The second inequality needs no such term because $\Lambda^{\frac\alpha2}\Psinu$ has zero mean. The endpoint $\alpha=\frac12$ is included by the Sobolev embedding $H^{1/4}(\T)\hookrightarrow L^4(\T)$ \cite{BenOh13}. At this endpoint, the exponent $1/(2\alpha)$ in the first inequality equals one, and the estimate remains valid. The bound on the evolution now reads
\begin{multline*}
\dfrac{d}{dt}\|\Lambda^{\frac\alpha2}\Psinu\|_{L^2(\T)}^2
+2\nu\|\Lambda^{\frac\alpha2+1}\Psinu\|_{L^2(\T)}^2
\leq C\|\Psinu\|_{L^2(\T)}^{2-\frac1\alpha}
\|\Lambda^{\frac\alpha2}\Psinu\|_{L^2(\T)}^{\frac1\alpha+\frac32}
\|\Lambda^{\frac\alpha2+1}\Psinu\|_{L^2(\T)}^{\frac12}\\
+C\|\Psinu\|_{L^2(\T)}^2
\|\Lambda^{\frac\alpha2}\Psinu\|_{L^2(\T)}^{\frac32}
\|\Lambda^{\frac\alpha2+1}\Psinu\|_{L^2(\T)}^{\frac12}.
\end{multline*}
Using the interpolation inequality
$$
\|\Lambda^{\frac\alpha2}\Psinu\|_{L^2(\T)}\leq\|\Psinu\|_{L^2(\T)}^{\frac{2}{\alpha+2}}\|\Lambda^{\frac\alpha2+1}\Psinu\|_{L^2(\T)}^{\frac{\alpha}{\alpha+2}},
$$
together with the Poincar{\'e} inequality on the $2\pi$-periodic torus,
$$
\|\Lambda^{\frac\alpha2}\Psinu\|_{L^2(\T)}\leq\|\Lambda^{\frac\alpha2+1}\Psinu\|_{L^2(\T)},
$$
we bound the first and second terms, respectively, and obtain
$$
\dfrac{d}{dt}\|\Lambda^{\frac\alpha2}\Psinu\|_{L^2(\T)}^2+2\nu\|\Lambda^{\frac\alpha2+1}\Psinu\|_{L^2(\T)}^2\leq C\|\Psinu\|_{L^2(\T)}^2\|\Lambda^{\frac\alpha2}\Psinu\|_{L^2(\T)}\|\Lambda^{\frac\alpha2+1}\Psinu\|_{L^2(\T)}.
$$
Then, using Young's inequality
$$
2ab\leq\nu^{-1}a^2+\nu b^2,\qquad a=\frac{C}{2}\|\Psinu\|_{L^2(\T)}^2 \|\Lambda^{\frac\alpha2}\Psinu\|_{L^2(\T)}, \qquad b=\|\Lambda^{\frac\alpha2+1}\Psinu\|_{L^2(\T)},
$$
we end up with the bound on the evolution of the Sobolev norm,
$$
\dfrac{d}{dt}\|\Lambda^{\frac\alpha2}\Psinu\|_{L^2(\T)}^2+\nu\|\Lambda^{\frac\alpha2+1}\Psinu\|_{L^2(\T)}^2\leq\dfrac{C}{\nu}\|\Psinu\|_{L^2(\T)}^4\|\Lambda^{\frac\alpha2}\Psinu\|_{L^2(\T)}^2,
$$
where the constant $C$ may change from line to line. Using the decay of mass we conclude that
$$
\|\Lambda^{\frac\alpha2}\Psinu(t)\|_{L^2(\T)}^2\leq\|\Lambda^{\frac\alpha2}\psi_0\|_{L^2(\T)}^2e^{\frac{Ct}{\nu}\mathcal{M}_0^2}.
$$
Applying the integrating factor $\exp(-C_\alpha\mathcal M_0^2t/\nu)$ to the differential inequality and retaining the dissipation term proves \eqref{eq:GWPViscBounds}. Together with the mass identity, it yields
\[
\int_0^T\|\Psinu(t)\|_{L^\infty(\T)}^2\,dt\leq C_\alpha\left(\mathcal M_0T+\frac{\|\Lambda^{\frac \alpha 2}\psi_0\|^2_{L^2(\T)}}{\nu}e^{C_\alpha\mathcal M_0^2T/\nu}\right)
\]
for every finite $T$ in the interval of existence, by the embedding $H^{\alpha/2+1}(\T)\hookrightarrow L^\infty(\T)$. The zero mode is controlled by the mass.

To close the continuation argument in $H^s$, set $J^s=(1-\Delta)^{s/2}$. The periodic tame product estimate \cite{BenOhZha25} gives
\[
\big\||\Psinu|^2\Psinu\big\|_{H^s}\leq C_s\|\Psinu\|_{L^\infty}^2\|\Psinu\|_{H^s}.
\]
In the $H^s$ energy estimate the dispersive term is skew-adjoint and the viscous term dissipative, so
\[
\frac12\frac{d}{dt}\|\Psinu\|_{H^s}^2+\nu\|\partial_x\Psinu\|_{H^s}^2\leq C_s\|\Psinu\|_{L^\infty}^2\|\Psinu\|_{H^s}^2.
\]
Gronwall's inequality and the preceding time-integrated $L^\infty$ bound rule out divergence of the $H^s$ norm at any finite maximal existence time. They also give $\Psinu\in L^2([0,T];H^{s+1})$, using the mass bound to control the zero mode. The solution therefore extends globally. The estimates are locally uniform on bounded sets of initial data in $H^s$; iterating the local stability result proves continuous dependence on every finite interval. Applying the same argument at every higher Sobolev order proves persistence of smoothness and completes the proof.

\section{Temporal scaling of the mass error in the inviscid limit}
\label{app:PreBlowMassGap}

Let us present heuristic arguments leading to the power law scaling presented in \eqref{eq:ScalingMassError}. First of all, defining
$$
\chi_\nu(t,x)= \partial_\nu \Psinu(t,x),
$$
one has
$$
\lim_{\nu_1,\nu_2\to 0} (\nu_2-\nu_1)^{-2} \mathcal{G}_{\nu_1,\nu_2}(t) = \|\chi_0(t,\cdot) \|_{L^2_x}^2.
$$
The claimed temporal scaling \eqref{eq:ScalingMassError} is therefore directly linked to the behavior of the derivative of the solution in the $\nu$ direction. Differentiating the viscous dynamics \eqref{eq:ViscousMMT} with respect to viscosity leads to
\begin{equation}
\label{eq:DynChi}
\begin{cases}
& i \partial_t \chi_\nu = \Lambda^\alpha\chi_\nu+ i\nu \Delta \chi_\nu + i\Delta \Psinu- 2\chi_\nu |\Psinu|^2 - \overline{\chi_\nu} (\Psinu)^2,\\
&\chi_\nu(0,x)=0.
\end{cases}
\end{equation}
Note that \eqref{eq:DynChi} is linear in $\chi_\nu$ and is forced by a source term $i\Delta \Psinu$. The source term $i\Delta\Psinu$ is the singular contribution in this equation. To extract its expected scaling heuristically, we use the collapse law observed in Fig.~\ref{fig:BlowupAnomDiss}. Let
\[
\tau=t^\star-t,\qquad A(t)=\|\psi(t)\|_{L^\infty_x}\sim \tau^{-1/2},
\]
and assume that the collapsing core is described by a single self-similar length scale $L(t)$. Balancing the time derivative, the cubic nonlinearity, and the fractional dispersion in the inviscid equation gives
\[
\tau^{-1}\sim A(t)^2 \sim L(t)^{-\alpha},\qquad\text{hence}\qquad L(t)\sim \tau^{1/\alpha}.
\]
For the case studied numerically, $\alpha=1/2$, this yields $L(t)\sim \tau^2$. The viscous sensitivity is forced by $\Delta\psi$. Assuming a self-similar collapse around the origin,
\[
\psi(t,x)\simeq A(t)Q\!\left(\frac{x}{L(t)}\right),
\]
one obtains the $L^2$ estimate
\[
\|\Delta\psi(t)\|_{L^2_x}\sim A(t)L(t)^{-2}L(t)^{1/2}= A(t)L(t)^{-3/2}.
\]
Using $A(t)\sim \tau^{-1/2}$ and $L(t)\sim \tau^{1/\alpha}$ therefore gives
\[
\|\Delta\psi(t)\|_{L^2_x}\sim \tau^{-1/2-\frac{3}{2\alpha}}.
\]
The Duhamel formula for the linearized equation then suggests, if the linearized propagator does not introduce a more singular scale than the forcing, that
\[
\|\chi_0(t)\|_{L^2_x}\sim \int^t \|\Delta\psi(s)\|_{L^2_x}\,ds\sim \tau^{\frac{1}{2}-\frac{3}{2\alpha}}.
\]
Consequently,
\[
\|\chi_0(t)\|_{L^2_x}^2\sim \tau^{1-\frac{3}{\alpha}}.
\]
For the simulations presented in the paper, $\alpha=1/2$, and this gives
\[
\|\chi_0(t)\|_{L^2_x}^2\sim (t^\star-t)^{-5}.
\]
Combining this estimate with
\[
\lim_{\nu_1,\nu_2\to 0}(\nu_2-\nu_1)^{-2} \mathcal{G}_{\nu_1,\nu_2}(t)=\|\chi_0(t)\|_{L^2_x}^2,
\]
one recovers the scaling observed numerically. This argument should be regarded as heuristic. The main uncontrolled point is the replacement of the full linearized evolution by the direct response to the forcing $i\Delta\psi$. The remaining linearized terms contain coefficients of size $|\psi|^2\sim \tau^{-1}$ and can affect prefactors, or possibly introduce additional algebraic corrections in a crude Gronwall estimate. The point of the argument is more modest: the observed exponent $5$ is the natural scaling produced by applying the viscous operator $\Delta$ to a collapsing fractional-NLS core with amplitude $A(t)\sim(t^\star-t)^{-1/2}$ and length scale $L(t)\sim(t^\star-t)^2$. Importantly, this scaling depends on the regularization. Following the same argument for the saturating nonlinearity, for which the source term is $|\Psisig|^6 \Psisig$, leads to the scaling $ (t^\star-t)^{\frac{1}\alpha-5}$. This gives an exponent $-3$ for $\alpha= \frac{1}{2}$.

\section{Control of fluctuations before blowup }
\label{app:AnomFluc}

Spontaneous stochasticity, and hence anomalous fluctuations, cannot happen while the solution of the problem remains unique. In this appendix, we prove bounds on the mass of fluctuations for both randomization mechanisms \eqref{eq:RandInitCond}.
\begin{proposition}[Pre-blowup control of fluctuations]
\label{prop:PreBlowFluctuations} Let $\alpha\geq1/2$, $s\geq\max\{4,\alpha/2\}$, and $\psi_0\in H^s(\T;\mathbb C)$, with $\mathcal M_0=\|\psi_0\|_{L^2}^2$. Let $\psi\in C([0,t^\star);H^s)$ be the solution of the inviscid equation \eqref{eq:MMT} with initial condition $\psi_0$, where $t^\star\in(0,\infty]$ is its maximal existence time. Consider the following two ensembles of solutions of the viscous equation \eqref{eq:ViscousMMT}:
\begin{enumerate}[label=\textnormal{(\roman*)}]
\item $\psi_{\nu,\mathbf i}^\omega$ has deterministic viscosity $\nu>0$ and initial condition \eqref{eq:RandInitCond}, with $\varepsilon=\nu$, $\lambda>0$, and independent phases $\theta_\omega^{\mathbf i}(k)$ uniformly distributed on $[0,2\pi]$.
\item $\psi_{\nu,\mathbf r}^\omega$ has deterministic initial condition $\psi_0$ and random viscosity $\nu(\omega)=\nu+\varrho_\nu^{\mathbf r}(\omega)$, where $\mathbb E_{\mathbf r}\varrho_\nu^{\mathbf r}=0$ and $|\varrho_\nu^{\mathbf r}|\leq\nu/2$ almost surely.
\end{enumerate}
For every $0<T<t^\star$, there exists $\nu_0=\nu_0(\alpha,s,\psi_0,T,\lambda)>0$ such that, for all $0<\nu\leq\nu_0$ and $0\leq t\leq T$, the fluctuation masses defined in \eqref{eq:DefFlucMass} satisfy
\begin{align}
\mathsf M_{\nu,\mathbf i}(t) &\leq 2\pi^2\nu^{2\lambda}\mathcal M_0 \exp\left[16\int_0^t\left(1+\|\psi(\tau)\|_{L^\infty}^2\right)d\tau\right], \label{eq:PropFlucInit}\\
\mathsf M_{\nu,\mathbf r}(t) &\leq C\nu^2\left[\int_0^t\left(1+\|\Delta\psi(\tau)\|_{L^2}^2\right)d\tau\right] \exp\left[C\int_0^t\left(1+\|\psi(\tau)\|_{L^\infty}^2\right)d\tau\right]. \label{eq:PropFlucVisc}
\end{align}
Here $C>0$ depends only on the fixed torus. The threshold $\nu_0$ is independent of the realizations and can be chosen uniformly over the viscosity distributions satisfying \textnormal{(ii)}. In particular, $\mathsf M_{\nu,\mathbf i}=O(\nu^{2\lambda})$ and $\mathsf M_{\nu,\mathbf r}=O(\nu^2)$ uniformly on $[0,T]$. These estimates make no assertion at or after $t^\star$ when it is finite.
\end{proposition}
We prove the two bounds separately below.

\subsection{The case of randomized initial conditions}
\label{app:AnomFlucViscInit}

Let $\psi_1=\psi_{\nu,\mathbf{i}}^{\omega_1}$ and $\psi_2=\psi_{\nu,\mathbf{i}}^{\omega_2}$ be solutions of \eqref{eq:ViscousMMT} with two independent realizations $\omega_1$ and $\omega_2$ of the initial condition $\psi_1(0)= \psi_0^{\nu,\omega_1}$ and $\psi_2(0)= \psi_0^{\nu,\omega_2}$. Because the solutions are independent and have the same average we have that
$$
\mathbb{E}^{(\omega_1, \omega_2)} \|\psi_1- \psi_2\|_{L^2}^2 = \mathbb{E}^{(\omega_1, \omega_2)} \|\varphi_1- \varphi_2 \|_{L^2}^2= 2\mathsf{M}_{\nu,\mathbf{i}}(t)
$$
Then, in order to write down a Gronwall bound on the norm of the difference, we use the factorization
$$
|\psi_1 |^2\psi_1- |\psi_2 |^2\psi_2=(\varphi_1 -\varphi_2)\left[ |\psi_1 |^2+ \overline{\psi_1}\psi_2 \right]+(\psi_2)^2 \overline{(\varphi_1- \varphi_2)}.
$$
This factorization leads us to the bound on the mass budget for the difference,
$$
\dfrac{d}{dt}\|\varphi_1- \varphi_2\|_{L^2}^2 \leq 2 \left( \|\psi_1\|_{L^\infty_x}+\|\psi_2\|_{L^\infty_x} \right)^2 \|\varphi_1- \varphi_2 \|_{L^2}^2\leq 4 \left( \|\psi_1\|_{L^\infty_x}^2+\|\psi_2\|_{L^\infty_x}^2 \right) \|\varphi_1- \varphi_2 \|_{L^2}^2.
$$
This bound can be integrated as
$$
\|\varphi_1- \varphi_2\|_{L^2}^2 \leq \|\varphi_1(0)- \varphi_2(0) \|_{L^2}^2 \exp\left(4 \int_0^t \left( \|\psi_1(s)\|^2_{L^\infty_x}+\|\psi_2(s)\|^2_{L^\infty_x} \right) ds \right)
$$
And in addition, because of our choice of perturbation of the initial condition we have the deterministic bound
$$
\|\varphi_1(0)- \varphi_2(0) \|_{L^2}^2 \leq (2\pi \nu^{\lambda})^2 \| \psi_0\|_{L^2_x}^2.
$$
All in all, using the independence of $\omega_1$ and $\omega_2$ and Cauchy-Schwarz, we end up with,
$$
\mathsf{M}_{\nu,\mathbf{i}}(t) \leq 2(\pi \nu^{\lambda})^2 \| \psi_0\|_{L^2_x}^2\mathbb{E}\left[ \exp\left(8\int_0^t \|\psi^\omega_\nu(s)\|^2_{L^\infty_x} ds \right)\right].
$$

The remaining expectation can be controlled by comparing directly with the smooth inviscid solution $\psi$. For smooth initial data and $T<t^\star$, the same Gronwall and bootstrap argument gives
$$
\|\psi_{\nu,\mathbf i}^\omega(t)-\psi_\nu(t)\|_{H^1}=O(\nu^\lambda),\qquad \|\psi_\nu(t)-\psi(t)\|_{H^1}=O(\nu),
$$
uniformly in $t\in[0,T]$ and in $\omega$, where $\psi_\nu$ has deterministic initial condition $\psi_0$. By Sobolev embedding, for sufficiently small $\nu$ this gives $\|\psi_{\nu,\mathbf i}^\omega(t)-\psi(t)\|_{L^\infty}\leq1$. Applying the triangle inequality in $L^\infty$ therefore yields
$$
\|\psi_{\nu,\mathbf i}^\omega(t)\|_{L^\infty}^2\leq 2\left(1+\|\psi(t)\|_{L^\infty}^2\right).
$$
Substituting this deterministic bound into the expectation above gives
\begin{equation}
\mathsf M_{\nu,\mathbf i}(t)\leq 2\pi^2\nu^{2\lambda}\mathcal{M}_0 \exp\left[16\int_0^t\left(1+\|\psi(s)\|_{L^\infty}^2\right)ds\right], \qquad 0\leq t\leq T<t^\star.
\end{equation}
Only the required smallness of $\nu$ depends on $T$ and the inviscid solution. At every fixed time before blowup, the integral is finite and the factor $\nu^{2\lambda}$ forces the fluctuations to vanish in the limit. At blowup, this control breaks down and the estimate no longer guarantees vanishing fluctuations.

\subsection{The case of randomized viscosity}
\label{app:RandomViscFluc}

Let $\psi_j=\psi_{\nu,\mathbf r}^{\omega_j}$, $j=1,2$, be two independent realizations with the same initial condition $\psi_0$ and viscosities $\nu_j=\nu(\omega_j)=\nu+\varrho_\nu^{\mathbf r}(\omega_j)$, as in \eqref{eq:RandInitCond}. Here $\nu$ is the deterministic average viscosity. For the randomization used in our simulations, $|\varrho_\nu^{\mathbf r}(\omega)|\leq\nu/2$. As in the previous section,
$$
\mathbb E_{\mathbf r}^{(\omega_1,\omega_2)}\|\psi_1-\psi_2\|_{L^2}^2=2\mathsf M_{\nu,\mathbf r}(t),\qquad \psi_1(0)-\psi_2(0)=0.
$$
The only difference is the additional term $(\nu_1-\nu_2)\Delta\psi_2$ in the equation for $\psi_1-\psi_2$. Using the same factorization of the cubic term as above, dropping the nonpositive dissipation term and applying Young's inequality gives
\begin{align*}
\frac{d}{dt}\|\psi_1-\psi_2\|_{L^2}^2\leq \left[1+4\left(\|\psi_1\|_{L^\infty}^2+\|\psi_2\|_{L^\infty}^2\right)\right]\|\psi_1-\psi_2\|_{L^2}^2 +|\nu_1-\nu_2|^2\|\Delta\psi_2\|_{L^2}^2.
\end{align*}
For smooth initial data and $T<t^\star$, the same comparison and bootstrap argument as above, now in $H^2$, gives $\|\psi_j(t)-\psi(t)\|_{H^2}\leq1$ uniformly in the realizations and in $t\in[0,T]$, for sufficiently small $\nu$. Here $\psi$ is the smooth inviscid solution. Thus $\|\psi_j\|_{L^\infty}^2\leq C(1+\|\psi\|_{L^\infty}^2)$ and $\|\Delta\psi_j\|_{L^2}^2\leq C(1+\|\Delta\psi\|_{L^2}^2)$. Since $|\nu_1-\nu_2|\leq\nu$, Gronwall's lemma, the identical initial conditions and averaging give
\begin{equation}
\label{eq:BoundMassFlucRandomVisc} \mathsf M_{\nu,\mathbf r}(t)\leq C\nu^2 \left[\int_0^t\left(1+\|\Delta\psi(s)\|_{L^2}^2\right)ds\right] \exp\left[C\int_0^t\left(1+\|\psi(s)\|_{L^\infty}^2\right)ds\right], \qquad 0\leq t\leq T<t^\star.
\end{equation}
The constant $C$ is independent of $\nu$ and $T$; only the required smallness of $\nu$ depends on $T$ and the inviscid solution. At every fixed time before blowup, the integrals are finite and the factor $\nu^2$ forces the mass of fluctuations to vanish. At blowup, this control by the smooth inviscid solution breaks down: the bound no longer guarantees vanishing fluctuations, leaving open the possibility of persistent randomness in the vanishing-viscosity limit.

\section{Numerical results concerning the saturating nonlinearity}
\label{app:SaturReg}

\begin{figure}[htb]
\centering
\includegraphics[width=0.95\linewidth]{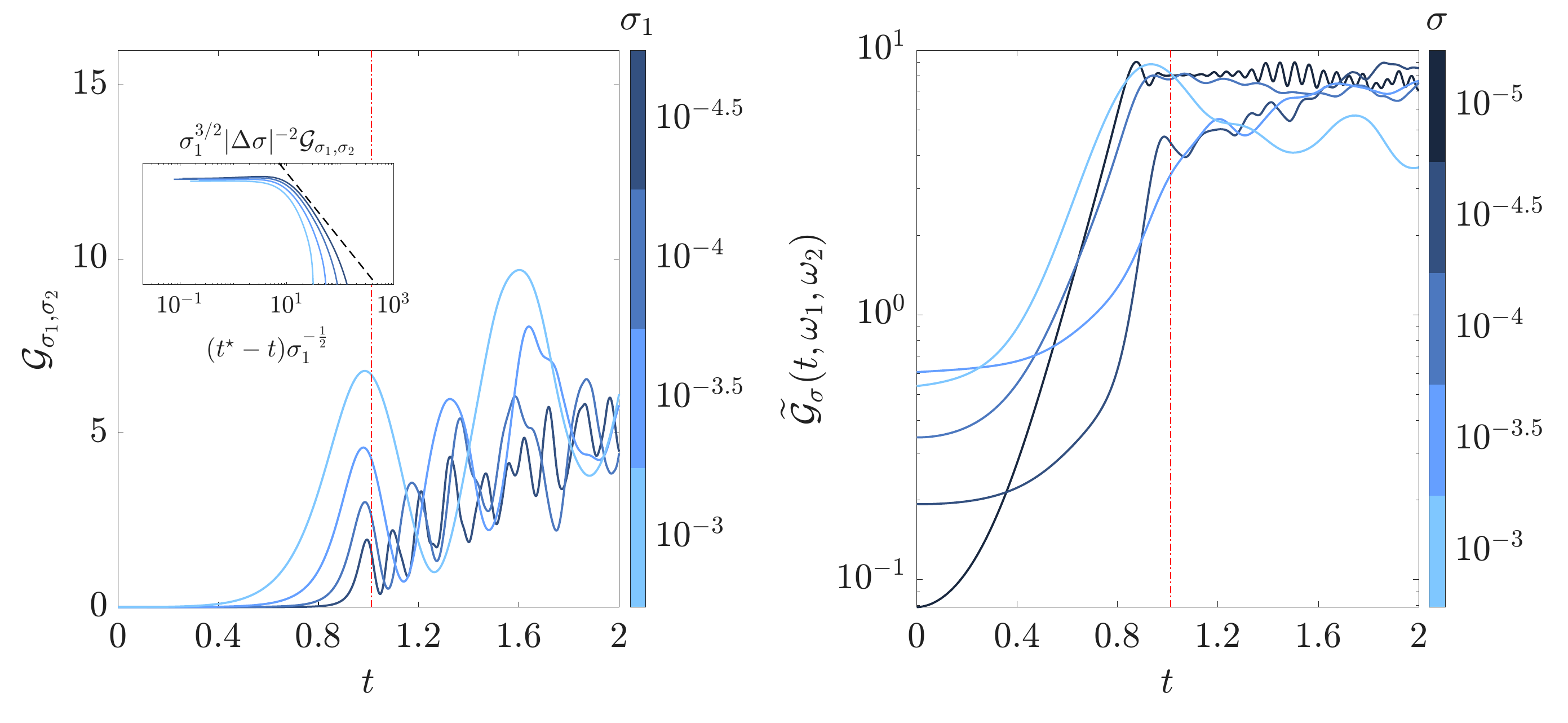}
\caption{Lack of deterministic selection for the saturating regularization \eqref{eq:SaturMMT}. Left: mass gap $ \mathcal{G}_{\sigma_1,\sigma_2}$ induced by changing the saturation parameter at fixed initial condition. The inset shows the same data rescaled by the arrest time \eqref{eq:TauSatur}, $\sigma^{3/2}|\Delta\sigma|^{-2}\mathcal{G}$ against $(t^\star-t)\sigma^{-1/2}$; colors match the main plot and the black dashed line shows the $-3$ power law. Right: mass gap $ \widetilde{\mathcal{G}}_{\sigma}(t,\omega_1,\omega_2)$ induced by two independent randomization of the initial condition. In both cases, the gap vanishes before $t^\star$ and remains finite after blowup as $\sigma\to 0$.}
\label{fig:LSPSatur}
\end{figure}

\begin{figure}
\centering
\includegraphics[width=0.95\linewidth]{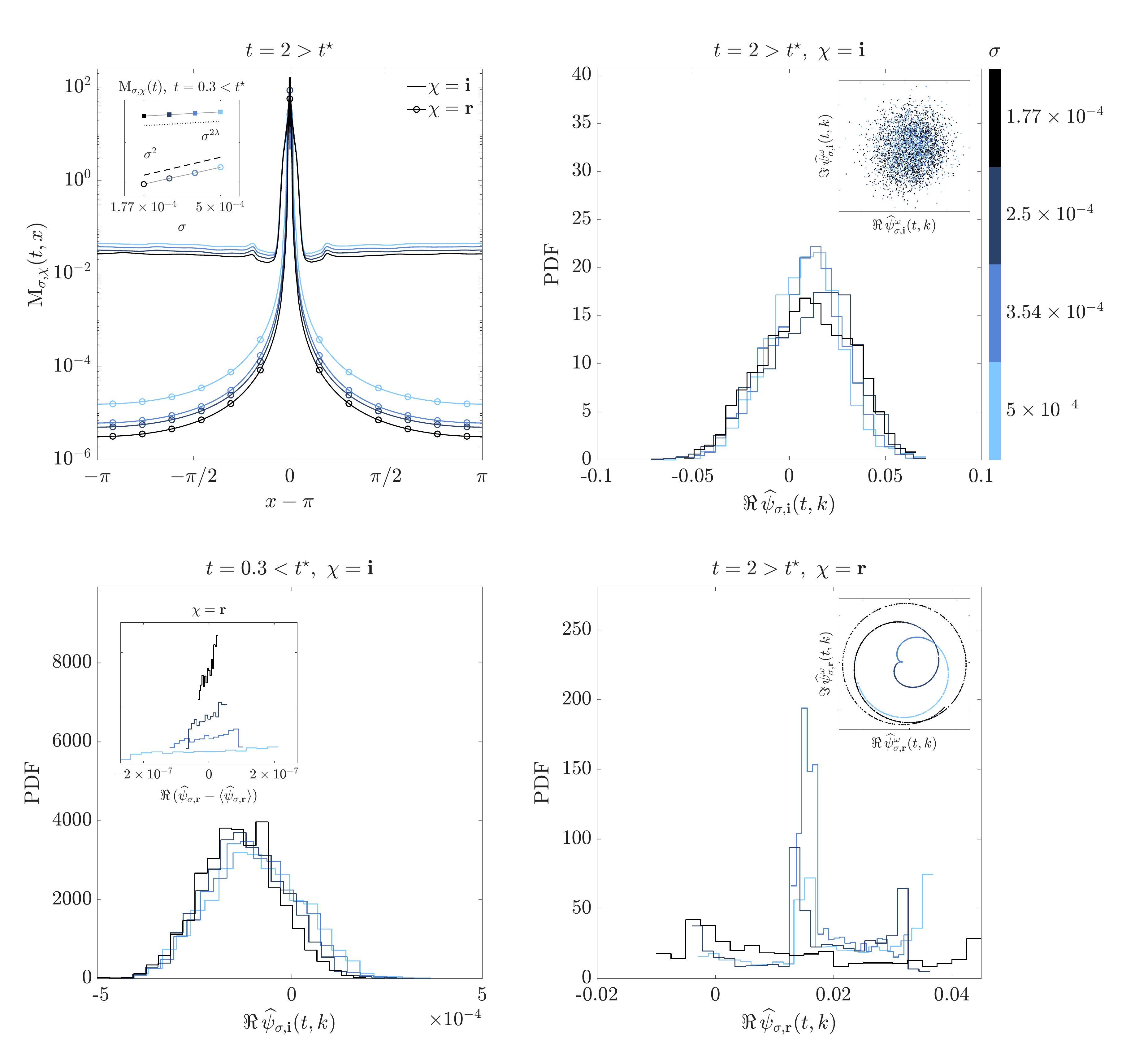}
\caption{Statistical description of the inviscid limit for the saturating regularization \eqref{eq:SaturMMT}, with the four saturation parameters $\sigma\in\{5,\,5/\sqrt2,\,5/2,\,5/(2\sqrt2)\}\times10^{-4}$ given in the colorbar. The four panels are the counterparts of those of Figure~\ref{fig:AnomFluc}. Upper left: local fluctuation mass $\mathsf{M}_{\sigma,\chi}(t,x)$ at $t=2>t^\star$, for randomized initial condition ($\chi=\mathbf{i}$, solid lines) and randomized saturation parameter ($\chi=\mathbf{r}$, circles); the inset shows the space-integrated fluctuation mass $\mathsf{M}_{\sigma,\chi}(t)$ at the pre-blowup time $t=0.3<t^\star$ against $\sigma$, together with the reference powers $\sigma^{2\lambda}$ (dotted) and $\sigma^{2}$ (dashed). Lower left: probability distribution of $\Re\widehat{\psi}_{\sigma,\mathbf{i}}(t,k)$ at $k=64$ before blowup, the inset showing the centered variable for $\chi=\mathbf{r}$. Upper right and lower right: the same distributions after blowup, for $\chi=\mathbf{i}$ and $\chi=\mathbf{r}$ respectively, the insets displaying $10^3$ independent samples of $\widehat{\psi}^\omega_{\sigma,\chi}(t,k)$ in the complex plane. In contrast with the viscous case, the accessible saturation parameters do not allow us to claim that the asymptotic regime $\sigma\to0$ has been reached.}
\label{fig:AnomFlucSatur}
\end{figure}

We gather here the numerical results concerning the saturating regularization \eqref{eq:SaturMMT}. The goal is to show that the lack of selection and the statistical description of the limit discussed in Sections~\ref{sec:LSP} and \ref{sec:StatisticalSelection} are not specific to the viscous regularization. The main difference is that \eqref{eq:SaturMMT} is conservative: for every fixed $\sigma>0$, one has $\M{\Psisig(t)}=\M{\psi_0}$ for all times. Thus, in contrast with the viscous case, post-blowup uncertainty is not accompanied by anomalous mass loss. We first test deterministic selection by comparing two solutions of \eqref{eq:SaturMMT} with the same initial condition and different saturation parameters by computing the mass gap $\mathcal{G}_{\sigma_1,\sigma_2}$. Before the first blowup time $t^\star$, this gap is expected to vanish as $\sigma_1,\sigma_2\to 0$, since the inviscid solution is smooth and unique. As sketched in Appendix~\ref{app:PreBlowMassGap}, we expect that before blowup,
\begin{equation}
\lim_{\sigma_1,\sigma_2 \to 0} |\sigma_1-\sigma_2|^{-2} \mathcal{G}_{\sigma_1,\sigma_2}(t) \underset{t\to t^\star}{\propto} (t^\star-t)^{\frac{1}{\alpha}-5}.
\end{equation}
For the case $\alpha=1/2$, this gives an exponent $-3$ which is consistent with the numerical data displayed in the inset of the left panel of Figure~\ref{fig:LSPSatur}. This power law holds only while the saturation is inactive. It ceases when the nonlinearity is throttled, that is when $\sigma|\Psisig|^4\sim1$. With the collapse law $A(t)\sim \tau^{-1/2}$ recalled in Appendix~\ref{app:PreBlowMassGap}, saturation is reached at the amplitude $A_\sigma\sim\sigma^{-1/4}$, hence at the time
\begin{equation}
\label{eq:TauSatur} \tau_\sigma \sim \sigma^{1/2},
\end{equation}
which plays for \eqref{eq:SaturMMT} the role that $\tau_\nu\sim\nu^{\alpha/(2-\alpha)}$ plays for \eqref{eq:ViscousMMT}. The gap therefore follows $|\Delta\sigma|^2\tau^{1/\alpha-5}$ down to $\tau\sim\tau_\sigma$, where the coarser of the two solutions stops collapsing and the growth halts. Removing both the size of the perturbation and this arrest scale, that is plotting
\[
\sigma^{3/2}|\Delta\sigma|^{-2}\,\mathcal{G}_{\sigma_1,\sigma_2} \qquad\text{against}\qquad (t^\star-t)\,\sigma^{-1/2},
\]
with $\sigma=\max(\sigma_1,\sigma_2)$, should therefore collapse the curves obtained at different saturation parameters onto a single master curve: flat below unity, and decaying with the $-3$ power above it. The prefactor $\sigma^{3/2}$ is the arrest time raised to the power of the temporal exponent, $\tau_\sigma^{3}\sim\sigma^{3/2}$. This is what the inset of the left panel of Fig.~\ref{fig:LSPSatur} displays, and the curves do collapse over the self-similar range, although less sharply than their viscous counterparts, the resolution we are able to reach with our numerical resources does not allow us to attain the asymptotic regime in a satisfactory way. After blowup, a finite limiting value indicates that the saturating regularization does not select a unique post-blowup continuation of \eqref{eq:MMT}. This behavior is observed in the left panel of Fig.~\ref{fig:LSPSatur}. We also test sensitivity to perturbations of the initial condition. For two independent realizations of the randomized initial condition \eqref{eq:RandomInit}, we compute $\widetilde{\mathcal{G}}_{\sigma}(t,\omega_1,\omega_2)$. The right panel of Fig.~\ref{fig:LSPSatur} shows that this gap also vanishes before blowup and remains finite after blowup in the vanishing-saturation limit. The arrest time \eqref{eq:TauSatur} is precisely the one entering the threshold derived in Section~\ref{sec:LCWRTIC}: a phase perturbation of size $\sigma^\lambda$ shifts the collapse time by $O(\sigma^\lambda)$, and the displacement measured in the inner collapse variables is $\sigma^\lambda/\tau_\sigma$, so that continuity is lost for $\lambda<\lambda_{\rm c}^{\sigma}=1/2$, against $\lambda_{\rm c}^{\nu}=\alpha/(2-\alpha)=1/3$ in the viscous case. The data reported here use $\lambda=1/4$, below both thresholds, so the post-blowup gap is expected to survive the limit for either regularization. This is what is observed: the gap at $t=2$ stays of order one across two decades of $\sigma$ while its value at initial time decreases as $\sigma^{2\lambda}=\sigma^{1/2}$. Thus the post-blowup dynamics is both nonselected by the saturation parameter and discontinuous with respect to vanishing perturbations of the initial condition.
We now turn to the statistical description of the inviscid limit. The notations of Section~\ref{sec:NoiseMechanisms} apply verbatim to the saturating regularization: the two randomizations \eqref{eq:RandInitCond} are used separately, with $\varepsilon=\sigma$, and the fluctuation mass $\mathsf{M}_{\sigma,\chi}$ is the one defined in \eqref{eq:DefFlucMass}. The only structural difference with the viscous case is that \eqref{eq:SaturMMT} conserves mass at every fixed $\sigma>0$, so that the fluctuations reported below are produced without any accompanying mass loss. The ensembles use $2\times10^3$ independent realizations at $N=2^{20}$ collocation points ($N=2^{21}$ for the smallest $\sigma$), for each of the four saturation parameters $\sigma\in\{5,\,5/\sqrt2,\,5/2,\,5/(2\sqrt2)\}\times10^{-4}$ and for each randomization $\chi$. The randomized initial data are perturbed at amplitude $\sigma^{\lambda}$ with $\lambda=1/4$, as for the mass gaps. Figure~\ref{fig:AnomFlucSatur} displays the same four diagnostics as Figure~\ref{fig:AnomFluc}, and the conclusions drawn for the viscous regularization in Section~\ref{sec:StatInvLimit} carry over.

Before blowup, the space-integrated fluctuation mass displayed in the inset of the upper left panel vanishes with $\sigma$, at rates compatible with the exponents $\sigma^{2\lambda}$ for $\chi=\mathbf{i}$ and $\sigma^{2}$ for $\chi=\mathbf{r}$ obtained in the viscous case in Proposition~\ref{prop:PreBlowFluctuations}. After blowup, the local fluctuation mass $\mathsf{M}_{\sigma,\chi}(t,x)$ remains finite as $\sigma\to0$ for both randomizations. It is strongly concentrated around the first collapse location $x=\pi$ while retaining a finite background away from it, and the symmetry about $x=\pi$ that each realization of \eqref{eq:RandInitCond} breaks is restored at the statistical level.

The distributions of $\Re\widehat{\psi}_{\sigma,\chi}(t,k)$ at $k=64$ display the same contrast between the two sources of randomness. For $\chi=\mathbf{i}$, the post-blowup distributions obtained at the four saturation parameters fall on top of one another, indicating convergence toward a non-Dirac limiting law, and the corresponding samples fill a two-dimensional region of the complex plane. For $\chi=\mathbf{r}$, the post-blowup distributions remain irregular and show no sign of convergence, while the samples lie on a one-dimensional curve, for the same reason as in the viscous case: a single real parameter is randomized.

Two reservations should be kept in mind. First, the accessible saturation parameters span less than a factor three, against a factor thirty in viscosity in Section~\ref{sec:StatInvLimit}, so that the vanishing-$\sigma$ trend is far less constrained. This is most visible in the pre-blowup distributions of the lower left panel: for $\chi=\mathbf{i}$ the expected narrowing is $\sigma^{\lambda}$, that is a factor $1.3$ over the available range, and is correspondingly difficult to resolve, whereas the much faster $\sigma$-dependence of the $\chi=\mathbf{r}$ case is clearly seen in the inset. Second, the solutions of \eqref{eq:SaturMMT} display long-lived, strongly fluctuating post-blowup structures, which makes ensemble averages less converged than in the dissipative case. The results of Figure~\ref{fig:AnomFlucSatur} should therefore be read as evidence that the same collapse-driven mechanism operates for the saturating regularization, rather than as a precise asymptotic characterization of the vanishing-saturation limit.
The viscous and saturating regularizations therefore lead to different mass balances in the inviscid limit: the former displays anomalous mass dissipation, whereas the latter remains conservative. Nevertheless, both regularizations fail to select a unique post-blowup continuation and both convert vanishing uncertainty into finite post-blowup fluctuations, with the same dependence on the source of randomness: convergence of the limiting distribution when the initial condition is randomized, and no sign of convergence when the regularization parameter is randomized instead. For the saturating nonlinearity this conclusion is, at this stage, mainly qualitative; reaching smaller values of $\sigma$ and improving statistical convergence will be necessary to establish the corresponding asymptotic regime more firmly. The observed behavior nevertheless indicates that the emergence of stochasticity is not a peculiarity of viscous dissipation, but is tied to wave collapse as a singular uncertainty-production mechanism.
\end{document}